\documentclass[acmlarge]{acmart}

\AtBeginDocument{%
  \providecommand\BibTeX{{%
    \normalfont B\kern-0.5em{\scshape i\kern-0.25em b}\kern-0.8em\TeX}}}

\usepackage{xspace}
\usepackage{soul}
\usepackage{tabularx}
\usepackage{caption}
\usepackage{subcaption}
\usepackage{adjustbox}
\usepackage{longtable}
\usepackage{array}
\usepackage{wrapfig,lipsum,booktabs}
\usepackage{textcomp}
\usepackage{multicol}
\usepackage{multirow}
\usepackage{comment}
\usepackage{amsmath}
\usepackage{enumitem,ragged2e}
\usepackage{xcolor,colortbl}

\setcopyright{acmlicensed}
\copyrightyear{2024}
\acmYear{2024}
\acmDOI{XXXXXXX.XXXXXXX}

\acmJournal{TOPS}
\acmVolume{0}
\acmNumber{0}
\acmArticle{1}
\acmMonth{1}

\newcommand{\V}{\mathcal{V}}
\newcommand{\E}{\mathcal{E}}
\newcommand{\G}{\mathcal{G}}
\newcommand{\N}{\mathcal{N}}

\definecolor{Gray}{gray}{0.85}
\definecolor{LightCyan}{rgb}{0.88,1,1}

\newif\ifcomments
\commentstrue

\begin{document}

\title{Use of Graph Neural Networks in Aiding Defensive Cyber Operations}

\author{Shaswata Mitra}
\email{sm3843@msstate.edu}
\orcid{}
\affiliation{%
  \institution{Mississippi State University}
  \city{Mississippi State}
  \state{MS}
  \country{USA}
  \postcode{39762}
}

\author{Trisha Chakraborty}
\email{tc2006@msstate.edu}
\orcid{}
\affiliation{%
  \institution{Mississippi State University}
  \city{Mississippi State}
  \state{MS}
  \country{USA}
  \postcode{39762}
}

\author{Subash Neupane}
\email{sn922@msstate.edu}
\orcid{}
\affiliation{%
  \institution{Mississippi State University}
  \city{Mississippi State}
  \state{MS}
  \country{USA}
  \postcode{39762}
}

\author{Aritran Piplai}
\email{apiplai@utep.edu}
\orcid{}
\affiliation{%
  \institution{University of Texas at El Paso}
  \city{El Paso}
  \state{TX}
  \country{USA}
  \postcode{79968}
}

\author{Sudip Mittal}
\email{mittal@cse.msstate.edu}
\orcid{}
\affiliation{%
  \institution{Mississippi State University}
  \city{Mississippi State}
  \state{MS}
  \country{USA}
  \postcode{39762}
}

\renewcommand{\shortauthors}{Mitra et al.}

\begin{abstract}
In an increasingly interconnected world, where information is the lifeblood of modern society, regular cyber-attacks sabotage the confidentiality, integrity, and availability of digital systems and information. Additionally, cyber-attacks differ depending on the objective and evolve rapidly to disguise defensive systems. However, a typical cyber-attack demonstrates a series of stages from attack initiation to final resolution, called an attack life cycle. These diverse characteristics and the relentless evolution of cyber attacks have led cyber defense to adopt modern approaches like Machine Learning to bolster defensive measures and break the attack life cycle. Among the adopted ML approaches, Graph Neural Networks have emerged as a promising approach for enhancing the effectiveness of defensive measures due to their ability to process and learn from heterogeneous cyber threat data. In this paper, we look into the application of GNNs in aiding to break each stage of one of the most renowned attack life cycles, the Lockheed Martin Cyber Kill Chain. We address each phase of CKC and discuss how GNNs contribute to preparing and preventing an attack from a defensive standpoint. Furthermore, We also discuss open research areas and further improvement scopes.  \label{section:abstract}
\end{abstract}




\maketitle
%35 pages
\section{Introduction}  \label{section:introduction}
% % What is Cybersecurity & Cyber operations
Cybersecurity is an ongoing practice dedicated to safeguarding systems and networks from cyberattacks that target unauthorized access, data alteration, information destruction, financial extortion, or disruption of regular business operations \cite{cisco_2022}. The significance of cybersecurity becomes more relevant with increased cyberattacks fueled by rapid digital expansion. Common avenues for cyberattacks encompass phishing, social engineering, password-related breaches, information misuse, man-in-the-middle attacks, denial-of-service (DoS) attacks, ransomware, and more. In response, cyber defense strategies incorporate a range of domains, network and perimeter security, endpoint security, application security, data security, identity \& access management, zero trust architecture, and more. These defense measures collectively aim to prevent attackers from achieving their goal through cyberspace, referred to as cyber defense operations.

% % Why ML and GNN
In an era characterized by unprecedented interconnectedness and technological advancement, the cybersecurity landscape faces an escalating challenge: the ceaseless barrage of cyber threats that target critical infrastructures, sensitive information, and even the very fabric of societies. As the digital landscape expands, so does the complexity of cyber attacks, necessitating a paradigm shift in defensive strategies. According to Gartner experts, of all cybersecurity breaches by 2024, 80\% will result from failures to prove the duty of due care~\cite{garner_cyber}. In response to constantly evolving attack patterns, enormous knowledge space, cost, and consistency, the marriage of machine learning (ML) and cybersecurity has yielded innovative solutions, propelling the evolution of defensive cyber operations to tackle unprecedented scenarios. Therefore, along with keeping the systems secure from possible threats, cyber defense operations now include getting ahead of the attacker to predict its possible next move based on the data pattern. However, it is impossible to effectively utilize ML techniques without a comprehensive, rich, and complete approach to the underlying data. Among the cutting-edge ML solutions, Graph Neural Networks (GNNs)~\cite{scarselli2008graph} have emerged as a promising contender, harnessing the power of neural networks to navigate and analyze intricate relationships within complex data structures. The introduction of GNN in defensive cyber operations introduces a new dimension of adaptive and intelligent defense mechanisms. Traditional cybersecurity approaches that rely on signature-based detection and rule-based techniques often struggle to keep pace with modern cyber threats' rapid mutation and polymorphism. In contrast, GNNs excel in capturing nuanced patterns and dependencies within diverse datasets, enabling them to uncover hidden insights that evade conventional methods. By viewing cyber threat data through a graph-based lens, GNNs inherently recognize the interconnectedness of entities, lending themselves to the intricate nature of cyber threat landscapes. 
% \hl{may be an example needed here rather than generic exmplanation- Subash} 

% % Research objective and section division
This survey paper aims to provide a comprehensive overview of the evolving cyber threat landscape where GNNs intersect with defensive cyber operations. To identify potential attack areas for a concrete countermeasure— we develop our taxonomy based on the Lockheed Martin cyber kill chain (CKC)~\cite{cyber_kill_chain} attack life cycle. The straightforward and simplistic approach is the primary reason behind appointing CKC as the skeleton behind our taxonomy. By exploration of relevant literature, this paper delves into the fundamental areas of cybersecurity following CKC and underpinning GNNs applications and the implications with opportunities for their integration into defensive strategies. By summarizing existing research and highlighting limitations, this survey aims to equip researchers with the capabilities and possible future research avenues in concretizing cyber defense operations through GNN. To the best of our knowledge, this survey is the first attempt that concentrates on the influence of GNN in overall cyber defense operations. In the following, we highlight our contributions in specific:  

% \hl{need to rephrase the last sentence}

% \hl{need to mention as to why CKC is chosen? Rn its only mentioned one time in contribution section. May be need few lines to elaborate on that.}

\begin{itemize}
    \item We demonstrate the application of GNNs in defensive cyber operations to detect and mitigate attacks by utilizing their knowledge propagation and learning capabilities.
    
    \item A comprehensive summary of the state-of-the-art research articles utilizing GNNs, grouped according to cover complete defensive cyber operations life-cycle through each cyber kill chain (CKC) phase. 

    \item We address the persisting challenges and discuss improvement scopes with future research directions in the employment of GNN in designing defensive cyber operation models.
    
\end{itemize}

The remainder of the paper is as follows: Section \ref{section:gnn} offers a foundational overview of Graph Neural Networks, elucidating their architecture, principles, and capabilities. Section \ref{section:taxonomy} delves into the intricate landscape of cyber attack and defensive cyber operations, providing the taxonomy of the integration of GNN. Section \ref{section:gnn_in_cyber} presents a comprehensive summary of existing literature following our taxonomy, showcasing the diverse applications of GNNs in cybersecurity contexts. Section \ref{section:discussion} critically examines the challenges and limitations of GNNs, addressing issues and potential improvement research directions. Finally, Section \ref{section:conclusion} concludes the survey by summarizing key findings, highlighting emergent trends, and proposing possible avenues for future research.

\section{Graph Neural Networks}  \label{section:gnn} %3 pages
 %Table for notation
%Define notations in text
%Preprocessing
%Message passing - aggregate and update
%node, graph, link level prediction
%variants of GNN -gcn, sage, ggnn, gat
%general GNN

The fundamental definition of graph neural network (GNN)~\cite{scarselli2008graph} denotes the adaptation of classical neural network (NN) framework on graph data to perform one or a combination of traditional machine learning tasks, such as classification, regression, or clustering. Data representation in graph form provides the means to capture complex relationships between entities and utilize these relations using two basic components -- nodes and edges. Existing vast literature in GNN has provided evidence for offering significant benefits to a broad range of domains that rely on artificial intelligence systems. Recently, systems based on variants of GNN have been well-explored on many tasks related to graph data, such as, social network \cite{Khaund2022Social} represents the interaction between social actors in the graph form where the vertices represent the social actors and the edges represent if there exists a relationship between two actors. In a citation network \cite{Cummings2022Structured}, two papers hold a relationship with each other via citations. In chemical sciences \cite{Fout2017Advances}, molecules are modeled as graphs, where the nodes represent protein molecules and edges, represent if their bio-activity exists in the chemical component and many more.

\subsection{Preliminaries}\label{gnn_preliminaries}
In this section, we define the graph notations and provide intuition to translate a real-world problem into a classification task. 
\smallskip

\noindent\textbf{Notations.} We denote a graph as $\G = (\V,\E)$, where $\V=\{v_1, v_2, ..., v_n\}$ is the set of nodes and $\E =\{e_1, e_2, \dots, e_m\} \\ \subseteq \V \times \V$ is set of edges between nodes. The \textit{neighborhood} of a node $v$ is denoted by set of nodes $\N(v) = \{u \in \V|(v,u) \in \E\}$. The node feature embeddings are represented by matrix $X = [x_1,\dots, x_n]^\top \in \mathbb{R}^{n\times d}$, where the $x_i \in \mathbb{R}^d$ is the $d$-dimensional feature vector of node $v_i$. Table~\ref{table:notation} provides a description of notations used following.

\begin{wraptable}{r}{8.5cm}
\begin{center}
\footnotesize
\begin{tabular} { | p {2.2 cm} | p {5.5 cm} | }
    \hline
    \rowcolor{LightCyan} 
    Notation & Description \\
    \hline
    $\G$ & Graph\\
    $\V$ & Set of nodes \\
    $\E$ & Set of edges \\
    $\N$(v) & Neighborhood of node v \\
    $n$ & Number of nodes \\
    $m$ & Number of edges \\
    deg($v$) & Degree of node $v$ \\
    $X = [x_1,\dots, x_n]^\top$ & Feature matrix \\
    $a_v^{(L)}$ & Message vector in $L$-th iteration\\
    $h_v^{(L)}$ &  Node representation of node $v$ in $L$-th iteration \\
    ${\mbox{AGGREGATE}^{(L)}(\cdot)}$ & Aggregation function in $L$-th iteration\\
    ${\mbox{UPDATE}^{(L)}(\cdot)}$ & Update function in $L$-th iteration \\
    \hline
\end{tabular}
\end{center}
\caption{Description of Notation.}
\label{table:notation}
\vspace{-8pt}
\vspace*{-5.5mm}
\end{wraptable}
\smallskip

\noindent\textbf{Problem Statement.} Recent literature has shown that GNNs can perform solve various challenging tasks on graph-structured data such as partitioning of graphs into meaningful subgraphs based on their structural properties~\cite{goh2003betweenness,bianchi2020spectral,muller2023graph,he2022sssnet} or generate new graphs that possess similar characteristics and properties from an input graph~\cite{wu2021self,xian2022generative,niu2020permutation}. In this survey, we begin by exploring classification tasks solved by GNN. Given a graph with nodes $v_i \in \V$ and edges $e_i \in \E$ representing entities and relationships, the task is to develop a GNN model that can effectively classify nodes, edges, or entire graphs into predefined $y$ classes, where $y \in \{y_1, y_2, \dots,y_c\}$ set of $c$ classes.

\begin{itemize}
    \item \textbf{Node-level Classification}. In node-level classification, the GNN model assigns each node to one or more predefined classes or predicts a continuous value associated with the node based on its node feature embedding from the neighboring nodes. 

    \item \textbf{Edge-level Classification.} In edge-level classification, the GNN should learn to classify edges into different categories or predict continuous values associated with them. The goal is to determine the nature of the relationship between pairs of connected nodes based on their neighboring node features and the edge itself.

    \item \textbf{Graph-level Classification.} In graph-level classification, the GNN should learn to classify entire graphs into specific categories. The GNN should effectively capture and aggregate information from all the nodes and edges in the graph to make predictions about the graph as a whole.
\end{itemize}

\subsection{Major GNN Frameworks} 
The ability of GNN to capture complex relationships originates from two prime components of the GNN framework: (1) aggregation and (2) update. Over multiple iterations, each node $v\in \V$ shares information containing the learned node feature with their neighborhood aggregates the node feature received from its neighborhood, and updates the node embedding using the learned/aggregated information~\cite{xu2019how, Ryoma2020Survey}. For a $L$-iteration, the general aggregation and updating function in $k\in L$ iteration is summarized as:
\vspace{-\baselineskip}

\begin{align}
     & a_v^{(k)} = {\mbox{AGGREGATE}}^{(k)}(h_u^{(k-1)}|u\in N(v)) &  (\forall k \in [L], v \in V), &  \\
     & h_v^{(k)} = {\mbox{UPDATE}}^{(k)}(h_v^{(k-1)},a_v^{(k)})    &  (\forall k \in [L], v \in V), 
\end{align}
\vspace{-\baselineskip}

where $a_v^{(k)}$ and $h_v^{(k)}$ represents message vector and the representation vector of node $v$ at the $k$-th iteration, respectively. At the first iteration of GNN, the initial node representation vector $h^{(0)} = X$. AGGREGATE($\cdot$) and UPDATE($\cdot$) are parameterized functions. The final node representation $H^{(L)}$ is fed into the classifier for identification, where $H^{(L)}=[h_v^{(L)}]$.
%%%%%%%%%%%%%%%%%%%%%%%%%%%%%%%%%%%%%%%%%%%%%%%%
\begin{figure*}
  \centering
  \includegraphics[width=1\textwidth, height = 3cm, trim = 1cm 15cm 2cm 10cm, clip]{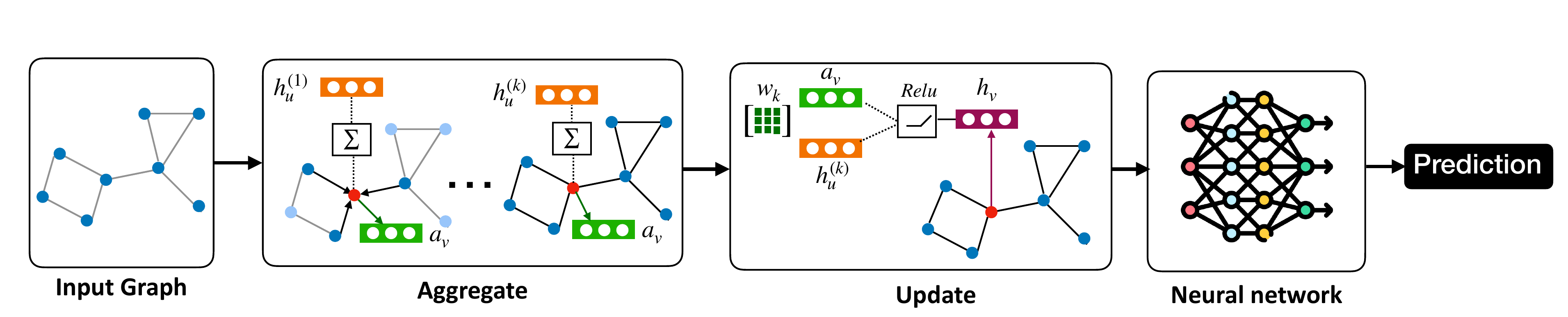}
  \caption{Illustration of GNN aggregate and update functions. The nodes $v$ of input graph $G$ is aggregates the embeddings $h_u^{(k)}$ over $k$ iteration from the neighbourhood $\N(v)$. Then, the node embedding of node $v$ is updated which is denoted by $a_v$. The final updated nodes are passed through a neural network for prediction. \textit{[Icons from~\cite{flaticon}]}} 
  \label{fig:gnn}
  \vspace{-3mm}
\end{figure*}
%%%%%%%%%%%%%%%%%%%%%%%%%%%%%%%%%%%%%%%%%%%%%%%%
With the advent of GNN frameworks in 2009~\cite{Scarselli2009GNN}, several GNN frameworks have been proposed, wherein its core, different choices of AGGREGATE($\cdot$) and UPDATE($\cdot$) functions lead to different variants of the GNN model. We briefly summarize the major GNN frameworks and showcase different aggregation and update functions related to the summarized GNN approaches described in Section \ref{section:gnn_in_cyber}.

\subsubsection{Graph Convolutional Network (GCN)}
Kipf et al.~\cite{kipf2016semi} proposed a framework that uses the convolution function on graph data to perform semi-supervised classification tasks. Each node in the graph learns the feature information from all its neighbors using the following aggregation and update function. 
\vspace{-2mm}
%%%%%%%%%%%%%%%%%%%%%%%%
\begin{equation}
\label{eq:gcn-agg}
    \mbox{AGGREGATE}^{(k)}(\{\{h_u^{(k-1}|u\in \N(v)\}\}) = \sum_{u\in \N(v)}\frac{h_u^{(k-1)}}{\sqrt{\mbox{deg}(v)\mbox{deg}(u)}} 
\end{equation}
\begin{equation}
\label{eq:gcn-up}
    \mbox{UPDATE}^{(k)}(h_v^{(k-1)},a_v^{(k)}) = \sigma(W^{(l)}a_v^{(k)}).
\end{equation}
%%%%%%%%%%%%%%%%%%%%%%%%

The aggregation process (equation~\ref{eq:gcn-agg}) for node $v$ is a function of feature information $h_v$ of all neighboring nodes $\N(v)$ from previous iteration $k-1$ iterations, where a non-parametric weight $\frac{1}{\sqrt{deg(v)deg(u)}}$ is assigned to any node $u$ and $v$. The aggregated message $a_v$ is used to update node $v$ (equation~\ref{eq:gcn-up}) learned representation using activation function $\sigma$, where $W(l)$ is a parameter matrix. While GCN provides a simplistic approach to learning graphical patterns, a major disadvantage of GCN is the overhead caused by computation for large graphs.

\subsubsection{GraphSAGE-mean}

Hamilton et al.~\cite{hamilton2017inductive} proposed a framework called GraphSAGE (SAmple and aggreGatE) to perform machine learning tasks on large graph-structured data (e.g. social network).  GraphSAGE uses a random sampling strategy to aggregate information from a node's neighbors to generate its embedding, unlike GCN~\cite{kipf2016semi} where all neighbors are considered.
\vspace{-1mm}
%%%%%%%%%%%%%%%%%%%%%%%%
\begin{equation}
\label{eq:sage-agg}
    \mbox{AGGREGATE}^{(k)}(\{\{h_u^{(k-1)}|u\in \N(v)\}\}) = \frac{1}{\mbox{deg}(v)}\sum_{u\in \N(v)}h_u^{(k-1)}
\end{equation}
\begin{equation}
\label{eq:sage-up}
    \mbox{UPDATE}^{(k)}(h_v^{(k-1)},a_v^{(k)}) = \sigma(W^{(l)}[h_v^{(k-1)},a_v^{(k)}]).
\end{equation}
%%%%%%%%%%%%%%%%%%%%%%%%

The aggregation process (equation~\ref{eq:sage-agg}) for node $v$ is a function of node embedding $h_v$ of all neighboring nodes $\N(v)$ from previous iteration $k-1$ iterations, where a non-parametric weight $\frac{1}{deg(v)}$. The update message $a_v$ (equation~\ref{eq:sage-up}) is used to update node $v$ learned representation using activation function $\sigma$, where a random sample of the node from $[h, a]$ is considered. The advantages of GraphSAGE include its ability to operate on large graphs with large numbers of nodes and edges. Due to the simplistic sampling technique that selects a subset of neighbors for each node, for a large graph, the process of aggregating is lightweight. Additionally, GraphSAGE can handle heterogeneous graphs where the dimensions of node features may differ.

\subsubsection{Graph Attention network (GAT)}

Veličković et al.~\cite{velickovic2018graph} proposed a graph attention network (GAT) that can learn the importance of each node when performing message passing. More specifically, when aggregating node information from neighboring nodes for each target node in the graph, the semantic similarity between the target node and each neighboring node will be considered by the multi-head attention mechanism, and important neighboring nodes will be assigned higher attention scores when performing the neighborhood aggregation.
\vspace{-1mm}
%%%%%%%%%%%%%%%%%%%%%%%%
\begin{equation}
\label{eq:gat-weight}
    \alpha_{uv}^{(l)} = \frac{\mbox{exp}(\mbox{LEAKYRELU}(a^{(l)T[W^{(l)}h_v^{(l-1)},W^{(l)}h_u^{(l-1)}]}))}{\sum_{u'\in \N(v)}{\mbox{exp}(\mbox{LEAKYRELU}(a^{(l)T}[W^{(l)}h_v^{(l-1)},W^{(l)}h_{u'}^{(l-1)}]))}}
\end{equation}
\begin{equation}
\label{eq:gat-agg}
    \mbox{AGGREGATE}^{(k)}(\{\{h_u^{(k-1}|u\in \N(v)\}\}) = \sum_{u\in \N(v)}\alpha_{vv}^{(l)}h_u^{(k-1)}
\end{equation}
\begin{equation}
\label{eq:gat-up}
    \mbox{UPDATE}^{(k)}(h_v^{(k-1)},a_v^{(k)}) = \sigma(W^{(l)}a_v^{(k)}).
\end{equation}
%%%%%%%%%%%%%%%%%%%%%%%%

Unlike GCNs, GATs use attention mechanisms to assign different weights to different nodes in the graph when aggregating information from neighbors. This allows GATs to learn more fine-grained representations that capture the relative importance of different nodes for each node in the graph.
Overall, GATs are a more powerful and flexible approach to node-level representation learning on graphs, but they can also be more computationally expensive than GCNs. The choice between GAT and GCN depends on the specific application and the computational resources available.

\subsubsection{Gated Graph Neural Network (GGNN)}

GGNN is a variant of recurrent GNN, which aims to learn node representations with recurrent neural architectures. Li et al.~\cite{Li2015GGNN} assumes that the node in a graph constantly exchanges information/message with its neighbors until a stable equilibrium is reached.
\vspace{-1mm}
%%%%%%%%%%%%%%%%%%%%%%%%
\begin{equation}
\label{eq:ggnn-agg}
    \mbox{AGGREGATE}^{(k)}(\{\{h_u^{(k-1}|u\in \N(v)\}\}) = A_i^T[h_1^{(k-1)}, \dots, h_n^{(k-1)}]^T
\end{equation}
\begin{equation}
\label{eq:ggnn-up}
    \mbox{UPDATE}^{(k)}(h_v^{(k-1)},a_v^{(k)}) = \mbox{GRU}(a_i^{(k)},h_i^{(k-1)}).
\end{equation}
%%%%%%%%%%%%%%%%%%%%%%%%

The aggregation process (equation~\ref{eq:ggnn-agg}) for node $v$ is a function of node embedding $h_v$ of all neighboring nodes $\N(v)$ from previous iteration $k-1$ iterations, where $\alpha_{uv}$ is the attention value for node $u$ and $v$. The update message $a_v$ (equation~\ref{eq:ggnn-up}) is used to update node $v$ learned representation using activation function $\sigma$. 
Unlike GCNs, GGNNs can handle dynamic graphs where the structure of the graph changes over time assisted by Gated Recurrent Units (GRU) that change in the graph structure over time.
GGNNs are more interpretable due to gates control the flow of information into and out of each node, allowing for a more fine-grained understanding of how information is being processed. GGNNs are a more powerful and flexible approach to node-level representation learning on dynamic graphs with longer-range dependencies, but they can also be more computationally expensive than GCNs. The choice between GGNN and GCN depends on the specific application and the characteristics of the graph data.

\section{Survey Overview and Taxonomy} \label{section:taxonomy} %3 pages
% Note: 
% Add the model citations.

% Include all aspects covering the point

% 1. Define Cyber operations with definitions and justify demarcation
% 2. Stages of cyber operations
% 3. GNN use in the stages
% 4. Add CKC inside the subsections

%What is cyber operation, attack life-cycle, life-cycle models requirement
The general definition of \textit{cyber operations} refers to the employment of cyberspace capabilities by an entity or organization towards a specific objective~\cite{nistco}. For the scope of this survey, we will refer to cyber operations from a defensive objective. Furthermore, a cyber attack comprises a series of steps by the attacker, leading from attack inception to the intended resolution, referred to as \textit{attack life-cycle}. Researchers have proposed several attack life-cycles ~\cite{mitre_attack_navigator, cyber_kill_chain, caltagirone2013diamond, nist_cybersecurity_framework, ibm_cybersattack_lifecycle, logrhythm_threat_lifecycle} to analyze the attack characteristics and design better cyber defense strategy. Among all these proposed attack life cycles, the diamond model~\cite{caltagirone2013diamond}, MITRE ATT\&CK~\cite{mitre_attack_navigator}, and Lockheed Martin cyber kill chain (CKC)~\cite{cyber_kill_chain} are most adopted by the industry. The \textit{diamond model} analyzes an attack by outlining the correlation between the attacker's motivations, victim, and infrastructures in the form of a diamond shape. Specifically, the model characterizes an attack through four quadrants: \textit{adversary, infrastructure, capabilities, and victim}. By examining the relationships between these quadrants, the diamond model helps understand the nature of the threat and develop strategies against it. \textit{MITRE ATT\&CK model} analyzes an attack by identifying detailed tactics, techniques, and procedures (TTPs) employed by the attacker throughout the entire attack life-cycle. It is organized as a matrix. Along the top row, it lists twelve tactical stages of an attack \textit{(initial access, execution, persistence, privilege escalation, defense evasion, credential access, discovery, lateral movement, collection, command \& control, exfiltration, and impact)} and possible techniques and procedures of each tactic in columns. It is comprehensive, covering a wide range of potential TTPs. Furthermore, the model gets regular updates to incorporate new attack TTPs as they are discovered. Lastly, Lockheed Martin's \textit{Cyber kill Chain} (CKC) describes the stages of a typical cyber attack as a linear chain of seven stages, namely: \textit{Reconnaissance, Weaponization, Delivery, Exploitation, Installation, Command \& Control, and Actions on Objectives}. This model provides a simplistic high-level view of the attack process, with a focus on the objective of the attacker in each stage. Due to the straightforward approach and wide acceptance throughout the industry, we will follow the CKC to analyze and develop our survey taxonomy.

On the other hand, traditional prevention techniques primarily rely on signature-based static approaches to disrupt these attack life cycles. Despite performing well with restricted cost, it fails to detect novel attacks and is easy to bypass with signature spoofing techniques. One of the leading reasons for the upward trend of ML in cybersecurity stems from this drawback. The pattern analysis capability and adaptation to behavioral changes lead to hybrid analysis that relies on static logic driven by dynamic behavioral data. To fuel this approach, data must contain complete, relevant, and rich contextual information to represent all potential outcomes. It should also be rich enough to cover all the details and relations between various working components like devices, applications, protocols, and network sensors to derive the right decision. Graph data structure perfectly fulfills these requirements. However, due to its heterogeneous nature, traditional ML models like CNN cannot utilize graph data for learning. Addressing this issue, GNN has emerged as a promising ML technique that can leverage the valuable threat information represented as graphical data to disrupt the attack life cycle.

% \vspace{-5mm}

% \includegraphics[width=\textwidth, height = 3cm, trim = 1cm 15cm 2cm 10cm, clip]

% Why ML and why GNN
% Talk more about graph data, logs, CTI, etc.
% To break the CKC, traditional prevention techniques mostly rely on signature-based static approaches. Even though it performs significantly well with restricted cost, but fails in detecting novel attacks and is easy to bypass with signature spoofing. 
% % \hl{the flow is breaking here...between two paragraphs...}
% One of the leading reasons for the upward trend of machine learning (ML) in cybersecurity stems from this drawback. The pattern analysis capability and adaptation to behavioral changes leads to dynamic analysis that doesn't just rely on static logic but also on the dynamic pattern of the data. To support such approach, data must contain complete, relevant, and rich context to represent the maximum number of potential outcomes. It should also be rich enough to contain all the details and relations between various working components like devices, applications, protocols, and network sensors to derive the right decision. Graph data structure perfectly fulfills such data requirements. However, due to the heterogeneous nature of graph data structure, traditional ML models like CNN cannot incorporate such learning. Hence, to tackle this problem, GNN has evolved as the promising ML approach capable of utilizing such rich information. 

\begin{figure*}[ht]
  \centering
  \includegraphics[width=\textwidth]{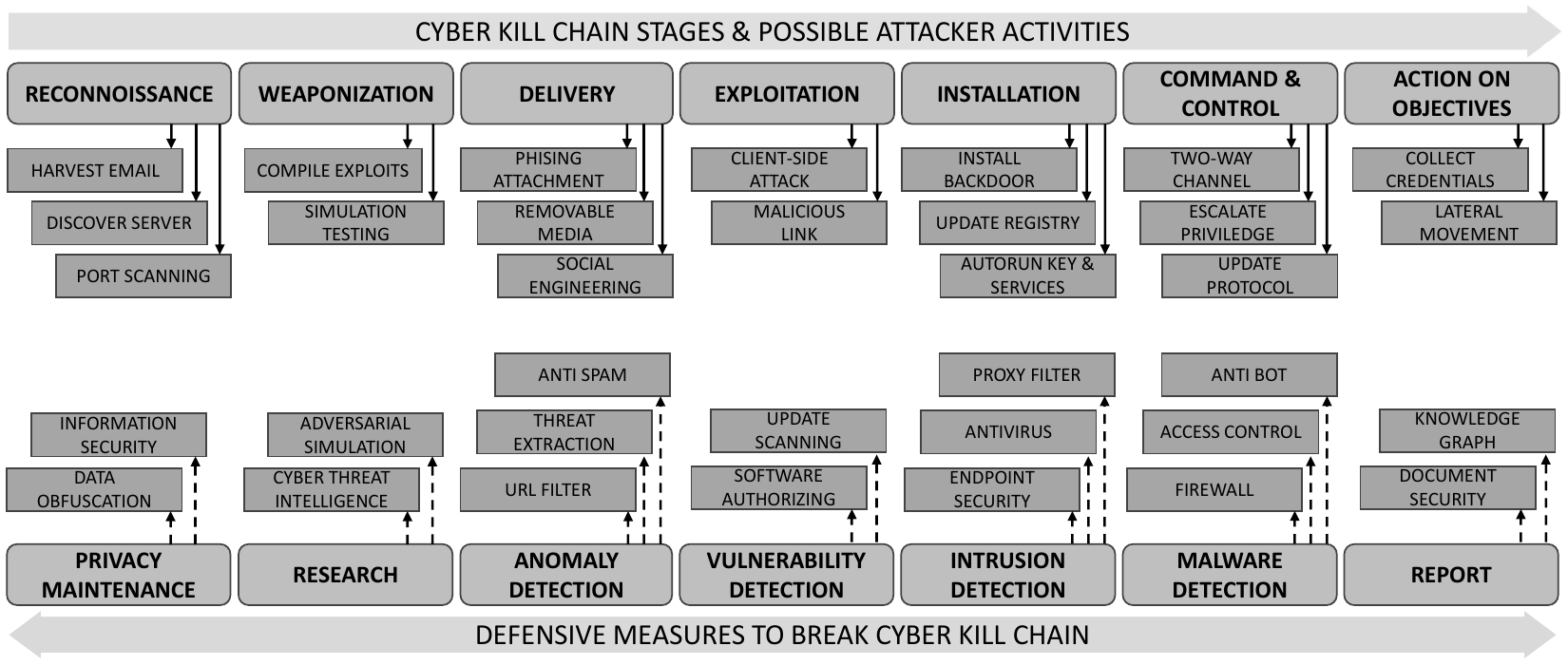}
  \caption{Overview of our proposed taxonomy. We considered seven phases of the cyber kill chain (CKC) [Reconnaissance, Weaponization, Delivery, Exploitation, Installation, Command \& Control, Actions on Objectives] with possible attacker activities. For prevention, we consider seven defensive phases [Privacy Maintenance, Research, Anomaly Detection, Vulnerability Detection, Intrusion Detection, Malware Detection, Report] with measures to counter and break the CKC from culmination.} 
  \label{Fig:ckc}
  \vspace{-5mm}
\end{figure*}

Following this, we will look into adversary activities through the lens of CKC and how GNN contributes from a defensive standpoint to counter each stage. We provide a summarized version of our taxonomy in Figure \ref{Fig:ckc}.
% \hl{may need to describe figure in few sentence..}
\vspace{-2mm}

% Diagram is not final,

\subsection{Reconnaissance} % [Done]
Reconnaissance is the first phase of the CKC, where the attacker identifies a victim and gathers information to discover vulnerabilities. It is accomplished by harvesting any information that can be useful to conduct an attack. For example, login credentials, user IDs, email addresses, system configuration, location, and others fall under the adversary's radar. To develop a concrete defense strategy, the defender should engage in continuous \textit{privacy maintenance} activities to prevent the adversary from gathering information. To ensure privacy, GNN can be used for link prediction, recommendation systems, and information embedding tasks to thwart adversaries from discovering sensitive information.

% \vspace{-2mm}
\subsection{Weaponization} % [Done]
Weaponization is the second stage, where attackers create/modify payload to take advantage of the discovered vulnerabilities in the target infrastructure to carry out an attack. This developed payload can be anything from a threat agent to a piece of malware, packaged to get delivered via any means to cause damage. From a counter-proactive standpoint, defenders engage in continuous \textit{research} by maintaining cyber threat intelligence (CTI) for possible attacks on the employed GNN models. By conducting such measures, defenders can maintain an upper hand in the ongoing cyber warfare and prevent significant damage. For example, simulated attacks on GNN utilizing known vulnerabilities can help discover and evaluate the effectiveness of existing security measures.

\vspace{-2mm}
\subsection{Delivery} % [Done]
Delivery is the third stage, referring to the transfer of weaponized payload to the target system. Attackers execute various delivery mechanisms like phishing, removable media, and social engineering, depending on the objective, target vulnerability, and desired stealth. From a defensive standpoint, organizations safeguard themselves against the delivery of malicious payloads by monitoring network or system anomalies. These measures include anti-spam, URL or email filtering, and more. Hence, \textit{anomaly detection} can be considered as a real-time malicious payload delivery prevention measure, aiming to detect suspicious items, network traffic, or observations. In such cases, GNNs play a vital role in detecting operational changes by learning network-system traffic activity graphs.

\vspace{-2mm}
\subsection{Exploitation} % [Done]
Exploitation is the fourth stage that involves the utilization of the vulnerability in the target system to initiate unauthorized access. During this, the attacker uses the delivered payload to exploit the vulnerability in the target system to execute malicious code or commands for initial foothold establishment. Some standard vulnerabilities arise from software bugs, misconfigurations, and other areas. As a preventative measure, \textit{vulnerability detection} involves identifying persisting loopholes to prevent exploitable scopes in source code, application patch updates, and security policies \& procedures. To fortify this, GNNs can learn the semantic flow of the application behavior and determine prevailing vulnerabilities before getting exploited by the delivered malicious payload.

\vspace{-2mm}
\subsection{Installation} % [Done]
Installation is the fifth phase, where the attacker establishes a persistence channel to solidify their foothold by installing threat actors into the compromised infrastructure. Attackers may modify registries, execute malicious files, and perform other actions depending on objectives and infrastructure. For prevention, defenders conduct \textit{intrusion detection} by reviewing signed certificates, application compile-time, analyzing privileges, and other relevant factors. Therefore, in a robust intrusion detection system, it is crucial to understand the system's topological structure. GNNs' ability to adapt and handle massive topological information makes them highly effective in identifying such malicious executions.

\vspace{-2mm}
\subsection{Command \& Control}
Command \& control (C2) is the sixth phase, referring to the communication and coordination infrastructure between attacker and victim's system or network to maintain persistent control by the attacker. Installed malware and escalated privilege allow attackers to issue commands, receive data, and sustain control within victim's environment. Hence, detecting and disrupting the C2 channel is essential in thwarting the ongoing attack. Security measures like \textit{malware detection} help intercept the C2 communication channel, thus limiting the attacker's control. Such detection requires a classification mechanism to identify system processes based on the execution pattern. GNN's innate ability to learn program flow aids in detecting and eradicating malware.

\vspace{-2mm}
\subsection{Actions on Objectives} 
Actions on objective is the final phase, during which an attacker accomplishes intended objectives such as data theft, sabotage, disruption, or other malicious goals. Attackers often erase or alter evidence of their presence and actions to maintain anonymity after goal accomplishment. Actions may be taken to impede security analysts from tracing, including deleting logs or manipulating timestamps. \textit{Reporting} attack TTPs is critical for analysis and further prevention. Activity logs, after-action reports, and cybersecurity knowledge graphs (CKGs) are shared globally to learn and identify attack patterns. In such scenarios, GNNs play a valuable role in learning from a vast amount of graphical data to support other security tools with improved knowledge and contextual understanding.
\newline

We define our taxonomy from a defensive cyber operations standpoint, where each category provides a direct countermeasure to the attack stage described above. Hence, our taxonomy is also divided into seven categories: \textit{privacy maintenance, research, anomaly detection, vulnerability detection, intrusion detection, malware detection, and report}. As mentioned, GNN has contributed significantly by utilizing its innate ability to process and learn from a tremendous amount of relational data in all these scenarios. This constitutes the rationale of our survey on the implementation of GNN in aiding cyber defense. Our taxonomy is based on the applications of GNN to break down each phase of the CKC. In the next section \ref{section:gnn_in_cyber}, we will address the abilities and drawbacks of existing research utilizing GNN to counter each phase of the CKC following the taxonomy. Later in Section \ref{section:discussion}, we will discuss potential areas of improvement, covering the identified drawbacks and future research directions.

\section{GNN's use in Cyber} \label{section:gnn_in_cyber}  % 11 pages % GNN Implementation Cases
In this section, we will discuss each preventive measure against the cyber kill chain in detail. First, we will discuss how the measures are beneficial in countering the attack phase using GNN. Then, we will delve deep into discussing some major recent developments in the last decade with a summarized table associated with each phase. We also provide a complete research information Table \ref{table:research_info} containing each article with its respective category, employed GNN models, classification types, experimental datasets, research timeline, and accuracy.
% \trisha{This is somewhere we can carefully word "how we selected these papers to survey"... maybe say... "Recent developments between 2018-2022"...}

\subsection{Privacy Maintenance} \label{subsection:privacy_summary}
    In the context of ML, much like the stages of the CKC, vulnerabilities threaten the privacy and integrity of models. Recent study~\cite{gong2020survey,neupane2023impacts} shows that ML models may suffer from the potential risk of leaking private information, are vulnerable to adversarial attacks~\cite{patterson2022white}, can inherit and magnify bias from training data, and lack interpretability~\cite{neupane2022explainable}. %which have a risk of causing unintentional harm to the users. 
% \st{A guideline of trustworthy AI systems has been proposed by the European Union}~\cite{trustworthy-ai} \st{indicates that trustworthy AI should obey the following four ethical principles: Respect for human autonomy, prevention of harm (privacy), fairness, and explainability.} 
%\hl{use kill chain oriented intro} 
In this section, we primarily emphasize the principle of privacy among other issues listed above. To safeguard privacy, researchers have developed privacy-preserving GNNs. These GNNs can be broadly categorized into four types: \emph{adversarial privacy preserving}, \emph{federated learning}, \emph{split learning}, and \emph{differential privacy}. Table \ref{table:privacy_preserving_gnns} provides an overview of the various privacy-preserving GNNs and their respective categorizations including other notable works. In the following, we will discuss state-of-the-art methods within each category. %\hl{need dataset citations for the summaries}
\begin{table*}[!htbp]
\centering
{
    \footnotesize
    {
    \def\arraystretch{1.3}
    \begin{tabular}{ p{3cm}|p{3cm}|p{5cm}|p{3.5cm}  }
    \rowcolor{lightgray!20!}
    %==========================================%
    \hline
    \textbf{Paper Title} & 
    \textbf{Focus/Objective} & 
    \textbf{Contributions} & 
    \textbf{Limitations}\\
    \hline
    %==========================================%
    \rowcolor{cyan!20!}
    \multicolumn{4}{c}{\textbf{Adversarial Privacy Preserving}} \tabularnewline
    \hline
   %==========================================%
    Adversarial privacy-preserving graph embedding against inference attack~\cite{li2020adversarial} &
    
    Integration of graph embedding and privacy protection into an end-to-end pipeline against inference attack.
 &
    
    \begin{minipage}[t]{\linewidth}
    \begin{itemize}[leftmargin=*]
    
        \item Graph embedding algorithm against inference attack.
        \item Introduced Privacy-Disentangling and Privacy-Purging.
    \end{itemize} 
    \vspace{1mm}
    \end{minipage} &
    \begin{minipage}[t]{\linewidth}
    \begin{itemize}[leftmargin=*]
    
        \item Full access of sensitive attribute might not be possible in real world.
    \end{itemize} 
    \vspace{1mm}
    \end{minipage} \\
  
    \hline

   %==========================================%
    Privacy-preserving representation learning on graphs: A mutual information perspective~\cite{wang2021privacy} &
    
    To learn node representations to achieve high performance for the primary learning task. &
    
    \begin{minipage}[t]{\linewidth}
    \begin{itemize}[leftmargin=*]
    
        \item Designs tractable algorithms to estimate intractable mutual information.
    \end{itemize} 
    \vspace{1mm}
    \end{minipage} &
    \begin{minipage}[t]{\linewidth}
    \begin{itemize}[leftmargin=*]
    
        \item Reduces the leakage of sensitive attributes but increases GNN training cost
    \end{itemize} 
    \vspace{1mm}
    \end{minipage} \\
  
    \hline 

 %==========================================%
    Information obfuscation of graph neural networks~\cite{liao2021information} &
    
    Formulate and address the problem of information obfuscation on graphs with GNNs.&
    
    \begin{minipage}[t]{\linewidth}
    \begin{itemize}[leftmargin=*]
    
        \item Creates a strong defense against information leakage while only suffering a marginal loss in task performance.

    \end{itemize} 
    \vspace{1mm}
    \end{minipage} &
    \begin{minipage}[t]{\linewidth}
    \begin{itemize}[leftmargin=*]
    
        \item To some context, use of custom perturbations on node for AT.
    \end{itemize} 
    \vspace{1mm}
    \end{minipage} \\
  
    \hline

    %==========================================%
    \rowcolor{cyan!20!}
     \multicolumn{4}{c}{\textbf{Federated Learning , Differential Privacy \& Split Learning }} \tabularnewline
    \hline
    
 Fedgnn: Federated graph neural network for privacy-preserving recommendation~\cite{wu2021fedgnn} & 
   
   Leverages FL and membership inference for privacy assurance. &
    \begin{minipage}[t]{\linewidth}
    \begin{itemize}[leftmargin=*]
        \item Designed novel federated framework for privacy-preserving GNN-based recommendation. 
        \item Introduced mechanism to protect model gradients in model training with local differential privacy.
    \end{itemize} 
    \vspace{1mm}
    \end{minipage}
    &
    
    \begin{minipage}[t]{\linewidth}
    \begin{itemize}[leftmargin=*]
        \item This method fails to solve social recommendations, and the clients models are not personalized

    \end{itemize} 
    \vspace{1mm}
    \end{minipage} \\
   
    %==========================================%
  \hline
    Towards representation identical privacy-preserving graph neural network via split learning~\cite{shan2021towards} &
    To address privacy issue of decentralized graphs using split learning and  horizontal FL.  & 
    \begin{minipage}[t]{\linewidth}
    \begin{itemize}[leftmargin=*]
        \item Generates the same node embedding as the centralized counterpart.
        \item Proposes a secure pooling mechanism instead of global pooling aggregator. 

    \end{itemize} 
    \vspace{1mm}
    \end{minipage} &
    \begin{minipage}[t]{\linewidth}
    \begin{itemize}[leftmargin=*]
        \item Do not use the formal notion of differential privacy and provide weaker privacy guarantees. 
    \end{itemize} 
    \vspace{1mm}
    \end{minipage} \\
    \hline
   
    %==========================================%
    \rowcolor{cyan!20!}
   \multicolumn{4}{c}{\textbf{Others}} \\
   \hline
    Learning privacy-preserving graph convolutional network with partially observed sensitive attributes~\cite{hu2022learning} 
    & Focuses on attribute inference attacks on GNNs.
    & 
    \begin{minipage}[t]{\linewidth}
    \begin{itemize}[leftmargin=*]
        
        \item Formulates a model to mitigate the individual privacy leakage using partially observed sensitive attribute. 

    \end{itemize} 
    \vspace{1mm}
    \end{minipage} &
    \begin{minipage}[t]{\linewidth}
    \begin{itemize}[leftmargin=*]
        \item Did not give differential privacy guarantees.
    \end{itemize} 
    \vspace{1mm}
    \end{minipage} \\
    \hline
    %==========================================%
      
      SecGNN: Privacy-preserving graph neural network training and inference as a cloud service ~\cite{wang2023secgnn}
    & Focuses on GNN training and inference services on cloud platforms. 
    & 
    \begin{minipage}[t]{\linewidth}
    \begin{itemize}[leftmargin=*]
        \item First system supporting privacy preserving GNN training and inference as a cloud service.
    \end{itemize} 
    \vspace{1mm}
    \end{minipage} 
    & 
    \begin{minipage}[t]{\linewidth}
    \begin{itemize}[leftmargin=*]
        \item Model might be weak for sophisticated adversaries.
    \end{itemize} 
    \vspace{1mm}
    \end{minipage} \\ 
    
    %==========================================%
    \hline
    \end{tabular}
    }
    \caption{
    A summary of the existing literature on privacy-preserving GNN with their scope, contribution, and limitations.}
    \label{table:privacy_preserving_gnns}
}
\vspace{-9mm}
\end{table*}

Significant efforts have been made in recent studies to enhance privacy protection and mitigate the risks of sensitive information leakage through modifications to the training process of GNNs. A notable example is the work of Li et al.~\cite{li2020adversarial}, where the concept of Adversarial Privacy Graph Embedding (APGE) to counter attribute inference attacks is introduced. The approach incorporates disentangling and purging mechanisms aimed at eliminating users' private information from the learned node representations. To achieve this, they integrate privacy-preserving regulation components into the loss function during the training phase. Experiments on Yale and Rochester datasets indicated the inclusion of disentangling and purging mechanisms brought significant gain in performance. Similarly, the authors in~\cite{wang2021privacy} present privacy protection utilizing adversarial privacy-preserving techniques. In the work, they propose a framework for privacy-preserving representation learning on graphs, adopting a perspective rooted in mutual information. Specifically, the framework is designed to address the concerns surrounding centralized training scenarios in GNNs and aims to prevent the inadvertent leakage of training data. To achieve this, they introduce bounding mechanisms that limit the exposure of sensitive information such as node features, node labels, and link status during GNN model training using three benchmark datasets including Cora~\cite{mccallum2000automating}, Citeseer~\cite{giles1998citeseer}, and Pubmed~\cite{sen2008collective}.
%\hl{Year:2021, Group: "Adversarial Privacy Preserving", Defense mechanism employed - "Attribute/Structure Reconstruction" Limitation:: Reduces the leakage of sensitive attributes but increases GNN training cost. .}~\cite{wang2021privacy}
%\hl{Year:2021, Group: "Adversarial Privacy Preserving", Defense mechanism employed -  information obfuscation-- "Attribute Reconstruction",  Limitation:: }~\cite{liao2021information} 2021
In a separate study~\cite{liao2021information}, researchers present a solution to combat attribute inference attacks on GNNs through an adversarial learning approach. The proposed Graph Adversarial Networks (GAL) framework approach revolves around a Wasserstein distance-based mini-max game between a desired graph feature encoder and a worst-case attacker for information obfuscation of sensitive attributes, empirically evaluated across 6 graph benchmark datasets including Citeseer~\cite{giles1998citeseer}, Pubmed~\cite{sen2008collective}, M9, ML-1M~\cite{konstan1997grouplens}, FB15K-237~\cite{toutanova2015representing}, and WN18RR. 
%\hl{ Year: 2021 Second paper is talking about p-p (Group: "Federeated learning" also some "Differential Privacy" by levereaging FL-- membership inference) in Federeated Recommender systems, limitation:: This method fails to solve social recommendations, and the clients models are not personalized.}~\cite{wu2021fedgnn} 

Another way to ensure privacy preservation is by utilizing Federated Learning (FL). FedGNN~\cite{wu2021fedgnn}, for instance, combines FL with GNNs in recommendation systems, aiming to safeguard users' privacy. This system captures complex user-item interactions by constructing local user-item graphs. Additionally, to address privacy concerns, the implementation of local differential privacy techniques helps mitigate privacy risks at the local level. The authors used six widely used benchmark datasets for recommendation, including MovieLens11 (100K,1M, and 10M) ~\cite{konstan1997grouplens}, Flixster~\cite{jamali2010matrix}, Douban~\cite{ma2011recommender}, and YahooMusic~\cite{dror2012yahoo}.
%\hl{Year 2021 : Group "Distrbuted System- Split learning" :: Limitation: Do not use the formal notion of Differential Privacy and provide weaker privacy guarantees}~\cite{shan2021towards}
In recent years, Split Learning (SL) has emerged as another prominent method and has captured the interest of researchers in the field of privacy maintenance tasks. SL involves partitioning a complete model into multiple sections. Building upon this concept,~\cite{shan2021towards} explores the integration of GNNs with SL to protect model privacy, specifically in node-level tasks, in scenarios where data is distributed horizontally across multiple silos and privacy is guaranteed via a secure min-max pooling aggregation mechanism. Similar to~\cite{wang2021privacy}, the authors also utilize the three same datasets for evaluation. 

%\hl{year: 2022: Limitation::Did not give differential privacy guarantees}~\cite{hu2022learning} 2022

In another line of work, the authors in~\cite{hu2022learning} delve into the realm of privacy-preserving GNNs, with a specific focus on handling partially observed sensitive attributes. Their primary objective is to address the issue of individual privacy leakage among private users. To achieve this, the authors propose a method that disentangles the node features into distinct latent representations of sensitive and non-sensitive attributes by imposing orthogonality within a suitable space. The authors evaluated their model on five benchmark datasets including two datasets with social networks, i.e., (Pokec-z and Pokec-n)~\cite{takac2012data}, and three ethical datasets constructed in ~\cite{agarwal2021towards}, i.e., German credit, Recidivism, and Credit defaulter. 
%\hl{year: 2023: Limitation::Model might be weak for  sophisticated adversaries}~\cite{wang2023secgnn} 2023
More recently, researchers have been focusing on deploying GNNs in cloud infrastructure to facilitate scalable and efficient graph analysis while maintaining privacy. One notable example is SecGNN~\cite{wang2023secgnn}, which enables privacy-preserving GNN training and inference services in the cloud. Unlike many existing approaches that rely on federated learning to protect privacy, with the primary emphasis on target system models and private training, SecGNN operates in an outsourced scenario. This means that the owner of the graph data can send its encrypted graph data to the cloud for secure training. In addition, SecGNN also supports secure GNN inference on encrypted GNNs and inputs, ensuring end-to-end security in the process. We further discuss the persisting loopholes and improvement areas in Section \ref{subsection:privacy_discussion}. 

% In the following Section \ref{subsection:research_summary}, we address recent developments in adversarial research as a counter-proactive measure against weaponization.

%\cite{li2020adversarial} 2020~\cite{wu2021fedgnn} 2021~\cite{shan2021towards} 2021~\cite{wang2021privacy} 2021~\cite{liao2021information} 2021~\cite{hu2022learning} 2022~\cite{wang2023secgnn} 2023 
    
\subsection{Research} \label{subsection:research_summary}

In the weaponization phase, the adversary develops an attack aimed at exploiting the identified vulnerabilities in the target system. Therefore, to maintain an effective security posture in an organization, the defender's job is to conduct adversarial simulations to determine possible drawbacks in the existing defensive ML models. One defensive approach is research, which allows defenders to evaluate and identify loopholes in end-systems or adopted models. This can be done by performing regular attack simulations using GNN on security ML models based on the reported threat intelligence. Information cover from globally documented cybersecurity databases (e.g., CVE) or scholarly articles. Below we discuss the research landscape and provide a summary in Table \ref{table:adversarial_gnns}.

% \begin{figure*}[h]
%   \centering
%   \includegraphics[width=0.65\textwidth]{research_diag.pdf}
%   \vspace{-3mm}
%   \caption{Research as a proactive measure against weaponization to prepare against possible attack.} 
%   \label{Fig:privacy_diag}
%   \vspace{-3mm}
% \end{figure*}

\begin{table*}[!htbp]
% \centering
{
    \footnotesize
    {
    \def\arraystretch{1.3}
    \begin{tabular}{p{3cm}|p{3cm}|p{5cm}|p{3.5cm}}
    \rowcolor{lightgray!20!}
    %==========================================%
    \hline
    \textbf{Paper Title} & 
    \textbf{Focus/Objective} & 
    \textbf{Contributions} & 
    \textbf{Limitations}\\
     %==========================================%
    \hline
    \rowcolor{cyan!20!}
    \multicolumn{4}{|c|}{\textbf{Research on Adversarial Attacks}} \\
    %==========================================%
     
     \hline
     Adversarial Attack on Graph Structured Data~\cite{dai2018adversarial}
     & 
     Developed adversarial attack on GNN-based graph supervised classification algorithm that can learn attack policy.
     & 
     \begin{minipage}[t]{\linewidth}
     \begin{itemize}[leftmargin=*]
         \item Considered white box, practical black box, and black box attack settings.
         \item Demonstrated attack with hierarchical reinforcement learning, random sampling, gradient-based, and genetic algorithm.
     \end{itemize} 
     \vspace{1mm}
     \end{minipage}
     & 
     \begin{minipage}[t]{\linewidth}
     \begin{itemize}[leftmargin=*]
         \item Limited experiments on real-world datasets to prove robustness and transferability.
         \item Do not consider poisoning attacks.
     \end{itemize} 
     \vspace{1mm}
     \end{minipage} \\

     %==========================================%
     \hline
     Adversarial Attacks on Graph Neural Networks via Meta Learning~\cite{zugner2019adversarial}
     & 
     Developed a training time meta-learning poisoning attack on GNN for node classification tasks.
     & 
     \begin{minipage}[t]{\linewidth}
     \begin{itemize}[leftmargin=*]
         \item Developed training time attack on attributed graphs.
         \item Used meta-gradient to solve bilevel training-time attacks.
     \end{itemize} 
     \vspace{1mm}
     \end{minipage}
     & 
     \begin{minipage}[t]{\linewidth}
     \begin{itemize}[leftmargin=*]
         \item Did not provide the working mechanism of the attack.
         \item No defensive measure is provided against such an attack.
     \end{itemize} 
     \vspace{1mm}
     \end{minipage} \\

     %==========================================%
     \hline
     Exploratory Adversarial Attacks on Graph Neural Networks~\cite{lin2020:epo}
     & 
     Developed an exploratory adversarial attack method to boost the gradient-based attacks on GNN.
     & 
     \begin{minipage}[t]{\linewidth}
     \begin{itemize}[leftmargin=*]
         \item Developed an exploratory approach to boost the gradient-based adversarial attacks.
     \end{itemize} 
     \vspace{1mm}
     \end{minipage}
     & 
     \begin{minipage}[t]{\linewidth}
     \begin{itemize}[leftmargin=*]
         \item Does not provide any defensive measure against the proposed approach.
     \end{itemize} 
     \vspace{1mm}
     \end{minipage} \\

     %==========================================%
     \hline
     Adversarial Attacks on Link Prediction Algorithms Based on Graph Neural Networks~\cite{lin2020adversarial}
     & 
     Developed an adversarial attack framework that generates perturbed graphs while preserving the overall structure to fool GNN-based link prediction tasks.
     & 
     \begin{minipage}[t]{\linewidth}
     \begin{itemize}[leftmargin=*]
         \item Developed an adversarial attack to fool GNN on link-prediction tasks.
         \item Generate perturbed graphs while preserving the overall structure to influence targeted output errors.
     \end{itemize} 
     \vspace{1mm}
     \end{minipage}
     & 
     \begin{minipage}[t]{\linewidth}
     \begin{itemize}[leftmargin=*]
         \item Did not provide any defensive measure against the proposed attack.
     \end{itemize} 
     \vspace{1mm}
     \end{minipage} \\

     %==========================================%
     \hline
     Backdoor Attacks to Graph Neural Networks~\cite{zhang2021backdoor}
     & Developed a subgraph-based backdoor attack to GNN for graph classification tasks.
     & 
     \begin{minipage}[t]{\linewidth}
     \begin{itemize}[leftmargin=*]
         \item GNN classifier predicts a targeted label if trained on the backdoor dataset.
         \item The backdoor attack does not impact GNN's accuracy over clean testing graphs.
     \end{itemize} 
     \vspace{1mm}
     \end{minipage}
     & 
     \begin{minipage}[t]{\linewidth}
     \begin{itemize}[leftmargin=*]
         \item Explored randomized smoothing-based certified defense but failed to provide a concrete solution.
     \end{itemize} 
     \vspace{1mm}
     \end{minipage} \\

     %==========================================%
     \hline
     Semantics-preserving Reinforcement Learning Attack Against Graph Neural Networks for Malware Detection~\cite{zhang2022semantics}
     & 
     Developed reinforcement learning-based semantics preserving adversarial attack against black-box GNNs for malware detection.
     & 
     \begin{minipage}[t]{\linewidth}
     \begin{itemize}[leftmargin=*]
         \item Used RL to insert Nops to preserve the program CFG semantics and elude GNN-based malware detectors.
         \item The attack is designed considering a black-box scenario and is transferable to similar models that work on sequential data.
     \end{itemize} 
     \vspace{1mm}
     \end{minipage}
     & 
     \begin{minipage}[t]{\linewidth}
     \begin{itemize}[leftmargin=*]
         \item Did not provide any defensive measure against proposed attack approaches.
     \end{itemize} 
     \vspace{1mm}
     \end{minipage} \\

     %==========================================%
     \hline
     Hierarchical Adversarial Attacks Against Graph Neural Network Based IoT Network Intrusion Detection System~\cite{zhou2021hierarchical}
     & 
     Developed a hierarchical adversarial attack generation method targeting state-of-the-art GNN-based black-box NIDS systems in IoT.
     & 
     \begin{minipage}[t]{\linewidth}
     \begin{itemize}[leftmargin=*]
         \item Developed framework for level-aware black-box attack strategy to generate samples using shadow GNN models.
         \item Used saliency map to identify critical feature elements and RWR mechanism for hierarchical node selection.
     \end{itemize} 
     \vspace{1mm}
     \end{minipage} 
     & 
     \begin{minipage}[t]{\linewidth}
     \begin{itemize}[leftmargin=*]
         \item Do not provide any defensive measure against the proposed attack approach.
         \item The Proposed approach is limited to NIDS in the IoT domain.
     \end{itemize} 
     \vspace{1mm}
     \end{minipage} \\

    %==========================================%
    \hline
    \rowcolor{cyan!20!}
    \multicolumn{4}{|c|}{\textbf{Research on Adversarial Defense}} \\
    %==========================================%

    %==========================================%
     \hline
     Developing graphical detection techniques for maintaining state estimation integrity against FDIA in integrated electric cyber-physical system\cite{li2020developing}
     & 
     Developed an FDIA detection method using GNN, combined with a capsule network to detect tampered measurements and attack locations.
     & 
     \begin{minipage}[t]{\linewidth}
     \begin{itemize}[leftmargin=*]
         \item Used GNN to detect tampered measurements without external knowledge.
         \item Used Capsule scheme to detect precise attack location and adapt with frequently evolving network structure. 
     \end{itemize} 
     \vspace{1mm}
     \end{minipage}
     & 
     \begin{minipage}[t]{\linewidth}
     \begin{itemize}[leftmargin=*]
         \item The GN detection model cannot adapt to frequently changing graphs.
         \item More practical experimentation is needed for robust implementation. 
     \end{itemize} 
     \vspace{1mm}
     \end{minipage} \\

     %==========================================%
     \hline
     Graph Neural Networks Based Detection of Stealth False Data Injection Attacks in Smart Grids~\cite{boyaci2021graph}
     & 
     Developed generic, localized, and stealth FDIA generation methodology and a scalable real-time GNN-based detector model.
     & 
     \begin{minipage}[t]{\linewidth}
     \begin{itemize}[leftmargin=*]
         \item Developed a stealth FDIA methodology that can bypass BDD systems.
         \item Developed a stochastic gradient descent-based stealth FDIA detection system.
     \end{itemize} 
     \vspace{1mm}
     \end{minipage} 
     &
     \begin{minipage}[t]{\linewidth}
     \begin{itemize}[leftmargin=*]
         \item Experiment is conducted over a randomly generated synthesized dataset.
     \end{itemize} 
     \vspace{1mm}
     \end{minipage} \\

    %==========================================%
    \hline
    \end{tabular}
    }
    \caption{
    Summary of the existing literature on adversarial research on GNN with scope, contribution, and limitations.}
    \label{table:adversarial_gnns}
}
\end{table*}

A hierarchical reinforcement learning adversarial attack by modifying the graph combinatorial structure for graph- and node-level classification tasks was proposed by Dai et al.~\cite{dai2018adversarial}. The modification is done by sequentially adding or dropping nodes or edges in the target graph. The attack is applicable in different adversarial settings with diverse GNN models for inductive and transductive learning tasks (\cite{scarselli2008graph, dai2016discriminative, kipf2016semi, hamilton2017inductive}). The model achieved a 40\% to 60\% success rate for graph-level attacks using randomized synthetic data and node-level attacks using Citeseer~\cite{giles1998citeseer}, Cora~\cite{mccallum2000automating}, Pubmed~\cite{sen2008collective}, and Finance datasets. In defense, the authors proposed incorporating adversarial samples in training, supported by the 1\% decrease in attack accuracy.
% Zügner paper - ADVERSARIAL ATTACKS ON GRAPH NEURAL NETWORKS VIA META LEARNING - 2019 [Done]
Similarly, Zügner et al.~\cite{zugner2019adversarial} investigated training time attacks using surrogate model meta-gradients on GNN for node classification tasks. Meta-learning helps the process to be more time and data-efficient by finding suitable hyperparameters. Evaluation in a black-box setting over Citeseer~\cite{giles1998citeseer}, Cora-ML~\cite{mccallum2000automating}, and POLBLOGS datasets~\cite{adamic2005political} confirmed GNN perform worse than a simple baseline of up to 48\% with a 5\% change limit. 
% Lin Paper - Exploratory Adversarial Attacks on Graph Neural Networks - 2020 [Done]
However, gradient-based attacks are limited in achieving sub-optimal results. Lin et al.~\cite{lin2020:epo} proposed an exploratory adversarial attack (EpoAtk) on GNN to boost the gradient-based perturbations and sidestep the possible misinformation provided by the maximal gradient. Experiments on semi-supervised node classification tasks over modified Citeseer~\cite{giles1998citeseer}, Cora~\cite{mccallum2000automating}, and Cora-ML~\cite{bojchevski2017deep} datasets revealed a drop in GCN accuracy with fewer perturb edges, outperforming the state-of-the-art attacks.

% Lin paper - Adversarial Attacks on Link Prediction Algorithms Based on Graph Neural Networks - Oct, 2020 [Done]
For white-box adversarial attacks against link prediction algorithms, Lin et al.~\cite{lin2020adversarial} developed a transferable evasion attack on the greedy heuristic that exploits incremental computation against a state-of-the-art link prediction algorithm called SEAL. It uses degree distribution preservation on the adjacency and the node information matrix to ensure un-noticeable perturbations in the sub-graph by an un-noticeability threshold. Evaluations over US Air, NS~\cite{newman2006finding}, Celegans~\cite{watts1998collective}, and PB~\cite{ackland2005mapping} datasets justify an approximated accuracy of 90\% with defined constraints.
% Zhang Paper  - Backdoor Attacks to Graph Neural Networks - June, 2021 [Done]
On the other hand, for the semi-black box graph classification setting, Zhang et al.~\cite{zhang2021backdoor} developed a sub-graph-based backdoor attack. It injects a subset of polluted sub-graphs in the training set to influence GNN to draw a correlation between the target and the trigger and predict the target label. Experiments on Bitcoin~\cite{weber2019anti}, Twitter~\cite{wang2017sybilscar}, and COLLAB~\cite{yanardag2015deep} datasets concluded its high success rate. Additionally in defense, the authors demonstrated works on certified defense using randomized sub-sampling. However, the findings are not generic and have a minor preventative impact, indicating further research requirements.

% Zhang Paper - Semantics-preserving Reinforcement Learning Attack Against Graph Neural Networks for Malware Detection - 2022 [Done]
To elude black-box malware detectors, Zhang et al.~\cite{zhang2022semantics} developed a semantic preserving reinforcement learning-based (SRL) attack. To maintain malware integrity, the model iteratively generates adversarial samples by sequentially injecting semantic Nops in the control flow graph (CFG). According to the results over labeled datasets from VXHeavens~\cite{tan2016artificial}, VirusShare~\cite{virusshare}, and VirusTotal~\cite{virustotal}, SRL was able to achieve 100\% evasion accuracy. 
% Zhou Paper - Hierarchical Adversarial Attacks Against Graph Neural Network Based IoT Network Intrusion Detection System - 2021 [Done]
Again, to attack GNN-based limited-budget IoT NIDS systems, Zhou et al.~\cite{zhou2021hierarchical} developed a Hierarchical Adversarial Attack (HAA) generation method. A shadow GNN model based on the saliency map technique generates adversarial examples by identifying and modifying critical node features with minimal perturbations. Random Walk with Restart mechanism (RWR) determines hierarchical vulnerable nodes based on the structural features and overall loss. Finally, to alter the classification labels, an adversarial shadow GNN disguises malicious packets as regular or vice versa. Experiments with UNSW-SOSR2019~\cite{hamza2019detecting} dataset over three baseline approaches proved degrading accuracy by more than 30\% against GCN and JK-Net.

% Li DBF FDI Paper - Developing graphical detection techniques for maintaining state estimation integrity against false data injection attack in integrated electric cyber-physical system - 2020 [Done]
As defensive research, to prevent false data injection attacks (FDIAs) and identify attack locations in electrical power networks, Li et al.~\cite{li2020developing} proposed a capsule scheme with GNN. The capsule scheme was combined with a dynamic routing mechanism to make GNN flexible to power topology changes and precisely identify attack locations. Experiments over different models with a consistent precision of 0.9939 and recall of 0.98231 on IEEE 30~\cite{ieee30} and 0.97131, and 0.97643 on IEEE 118-bus systems~\cite{ieee118}, showcased its robustness. 
% Boyaci Paper - Graph Neural Networks Based Detection of Stealth False Data Injection Attacks in Smart Grids - Aug 2021 [Done]
Similarly, Boyaci et al.~\cite{boyaci2021graph} proposed another FDIA detector using GNN for early warnings in smart grid systems. The model was designed to handle dynamic measurement data by learning the underlying topological structure alongside data patterns. Additionally, the authors developed a generic, localized, and stealth FDIA generation methodology and publicly accessible datasets for additional research. Experiments were conducted over synthetic data using Pandapower~\cite{thurner2018pandapower} over ERCOT's load profiles for IEEE 14, 118, and 300 buses ~\cite{ieee14, ieee118, ieee300}, and the proposed model outperformed the benchmark CNN models by 3.14\%, 4.25\%, and 4.41\%. In Section \ref{subsection:research_discussion}, we further discuss persistent loopholes and areas for improvement. \\

% In the next Section \ref{subsection:anomaly_summary}, we cover recent developments in anomaly detection to prevent the delivery of malicious payloads.
\vspace{-5mm}

\subsection{Anomaly Detection} \label{subsection:anomaly_summary}
    Anomaly detection refers to identifying suspicious items, events, or observations that exhibit significantly distinctive characteristics from standard patterns or behaviors. In cybersecurity, anomalies can occur due to numerous factors, such as initiation/delivery of malware attacks, network intrusions, social engineering, or insider threats. Detecting these anomalies is critical for identifying and preventing the delivery of potential malicious payloads. Furthermore, anomaly detection can provide early warning signs of an attack, reduce false positives, and help to improve the overall security posture. Anomalies are of three types: \emph{point, contextual/conditional}, and \emph{collective}~\cite{wu2021graph}. In the context of GNN, point anomaly refers to node or edge classification, whereas contextual and collective anomalies are primarily graph classification. The potential of GNN in anomaly detection tasks has proven beneficial, which we elaborate on in the following paragraphs with a summary in Table \ref{table:anomaly_gnns}.

\begin{table*}[!htbp]
% \centering
{
    \footnotesize
    {
    \def\arraystretch{1.3}
    \begin{tabular}{ p{3cm}|p{3cm}|p{5cm}|p{3.5cm}  }
    \rowcolor{lightgray!20!}
    %==========================================%
    \hline
    \textbf{Paper Title} & 
    \textbf{Focus/Objective} & 
    \textbf{Contributions} & 
    \textbf{Limitations}\\
     %==========================================%
    \hline
    \rowcolor{cyan!20!}
    \multicolumn{4}{|c|}{\textbf{Point Anomaly}} \\
    %==========================================%

    %==========================================% 
     \hline
     Anomaly Detection using Graph Neural Networks~\cite{chaudhary2019anomaly} 
     & 
     Detecting spam, fake reviews, or malicious activities in email (Enron) and social networks (Twitter). 
     & 
     \begin{minipage}[t]{\linewidth}      \begin{itemize}[leftmargin=*]
        \item Developed a GNN model to detect anomalies in mail and social network graphs.
        \item GNN encoding achieved a classification accuracy of over 98\%.
     \end{itemize}       \vspace{1mm}      \end{minipage} 
     &
     \begin{minipage}[t]{\linewidth}      \begin{itemize}[leftmargin=*]
         \item Experiment was conducted over a limited data set.
         \item Architecture was designed to cover specific scenarios.
     \end{itemize}       \vspace{1mm}      \end{minipage} \\
     
    %==========================================%    
     \hline
     Relevance-Aware Anomalous Users Detection in Social Network via Graph Neural Network~\cite{li2021relevance} 
     & 
     Develop RAU-GNN to acquire fine-grained detection results to identify abnormal social users like bots, spammers, etc. 
     & 
     \begin{minipage}[t]{\linewidth}      \begin{itemize}[leftmargin=*]
         \item Fusion GCN layer is used for anomalous behavior classification.
         \item Incorporated a Multi-head Graph Attention Network (GAT) with GCN to detect camouflaging users.
     \end{itemize}       \vspace{1mm}      \end{minipage}
     & 
     \begin{minipage}[t]{\linewidth}      \begin{itemize}[leftmargin=*]
         \item GAT embeddings rely on GCN node representations.
         \item Multiple computation-heavy (GAT, GCN) layers might increase computation. 
     \end{itemize}       \vspace{1mm}      \end{minipage} \\

    %==========================================%
     \hline
     \rowcolor{cyan!20!}
     \multicolumn{4}{|c|}{\textbf{Contextual Anomalies}} \\

    %==========================================%
     \hline
     An Anomaly Event Detection Method Based on GNN Algorithm for Multi-data Sources~\cite{ji2021anomaly} 
     & 
     Detect anomalies in multi-source data in the secure critical infrastructure domain. 
     & 
     \begin{minipage}[t]{\linewidth}      \begin{itemize}[leftmargin=*]
         \item Developed Deep-GNN model to incorporate heterogeneous data set.
         \item PCA was used in data pre-processing to increase robustness.
     \end{itemize}       \vspace{1mm}      \end{minipage}
     & 
     \begin{minipage}[t]{\linewidth}      \begin{itemize}[leftmargin=*]
         \item GNN heavily relies on spectral clustering for feature extraction and data fusion.
     \end{itemize}       \vspace{1mm}      \end{minipage} \\
     
    %==========================================%
     \hline
     Spammer Detection Using Graph-level Classification Model of Graph Neural Network~\cite{song2021spammer}
     & 
     Developed a GNN model to detect spammers in social media.
     & 
     \begin{minipage}[t]{\linewidth}      \begin{itemize}[leftmargin=*]
         \item Converted user behavior graph to extract features for model training.
         \item Used Bayesian optimization framework for hyperparameters tuning.
         \item Used 10-fold cross-validation.
     \end{itemize}       \vspace{1mm}      \end{minipage} 
     & 
     \begin{minipage}[t]{\linewidth}      \begin{itemize}[leftmargin=*]
         \item Experimental setup is limited to a single dataset.
         \item Time variations can be incorporated to allow dynamic graph classification.
     \end{itemize}       \vspace{1mm}      \end{minipage} \\
 
    %==========================================%    
     % \hline
     % Graph Neural Networks for Anomaly Detection in Industrial Internet of Things~\cite{wu2021graph} 
     % & 
     % Detecting point, contextual, and collective anomaly in IIoT, using smart-transportation, smart-energy, and smart-factory industry case studies. 
     % & 
     % \begin{minipage}[t]{\linewidth}      \begin{itemize}[leftmargin=*]
     %     \item Developed GCN-based model on the Siamese network with k-nearest neighbor techniques to detect point anomaly.
     %     \item Developed a multi-layer Spatial-Temporal Graphic Neural Network (STGNN) for detecting contextual anomalies in smart energy.
     %     \item Developed an n-layer GCN model with a SoftMax to detect contextual anomaly.
     % \end{itemize}       \vspace{1mm}      \end{minipage}
     % & \begin{minipage}[t]{\linewidth}      \begin{itemize}[leftmargin=*]
     %     \item The paper only focused on three sectors (smart transportation, smart energy, and smart factory) of IIoT with limited improved aspects.
     %     \item The GNN model is heavily interdependent on the other ML models that are associated with it.
     % \end{itemize}       \vspace{1mm}      \end{minipage} \\
     
    %==========================================%
    \hline
    \rowcolor{cyan!20!}
    \multicolumn{4}{|c|}{\textbf{Collective Anomaly}} \\
    %==========================================%

    %==========================================%
     \hline
     One-class graph neural networks for anomaly detection in attributed networks~\cite{wang2021one} 
     & 
     Generate normal node embedding close to the center of the hypersphere for precise anomaly detection.
     & 
     \begin{minipage}[t]{\linewidth}      \begin{itemize}[leftmargin=*]
         \item OCGNN can generate more accurate embeddings in close proximity.
         \item Hypersphere learning allows the model to be robust from a usability standpoint.
     \end{itemize}       \vspace{1mm}      \end{minipage} 
     & 
     \begin{minipage}[t]{\linewidth}      
     \begin{itemize}[leftmargin=*]
         \item The model cannot handle large dynamic graphs.
     \end{itemize}       \vspace{1mm}      \end{minipage} \\

      %==========================================%    
     \hline
     One-class Temporal Graph Attention Neural Network for Dynamic Graph Anomaly Detection~\cite{huang2021one} 
     & 
     Generate natural node embedding close to the center of the hypersphere while incorporating dynamic graphs.
     & 
     \begin{minipage}[t]{\linewidth}      \begin{itemize}[leftmargin=*]
         \item TGAT is used to extract large dynamic graph representation.
         \item The approach is capable of edge classification and link prediction.
     \end{itemize}       \vspace{1mm}      \end{minipage}
     & 
     \begin{minipage}[t]{\linewidth}      
     \begin{itemize}[leftmargin=*]
         \item The indirect classification mechanism and requirement of anomaly-free training data might make the model scope limited.
     \end{itemize}       \vspace{1mm}      \end{minipage} \\

     %==========================================%    
     \hline
     Multi-layer Graph Neural Network-Based Random Anomalous Behavior Detection~\cite{shi2021multi} 
     & 
     Developed a random anomalous edges detection model for an abundant edge graph. 
     & 
     \begin{minipage}[t]{\linewidth}     \begin{itemize}[leftmargin=*]
         \item Combined the GCN and GAT to learn abstract node features from the differentiated local structure influence.
         \item The applicability of the model was tested with a diverse dataset.
     \end{itemize}       \vspace{1mm}      \end{minipage} 
     & 
     \begin{minipage}[t]{\linewidth}      \begin{itemize}[leftmargin=*]
         \item The model was more inclined towards random anomalous edge detection.
     \end{itemize}       \vspace{1mm}      \end{minipage} \\

      %==========================================%    
     \hline
     \rowcolor{cyan!20!}
     \multicolumn{4}{|c|}{\textbf{Others}}\\
 
    %==========================================%    
     \hline
     GNN-based Graph Anomaly Detection with Graph Anomaly Loss~\cite{zhao2020gnn} 
     &
     To represent anomaly-detectable nodes for better anomaly and dense block detection using improved GNN loss function. 
     &
     \begin{minipage}[t]{\linewidth}     \begin{itemize}[leftmargin=*]
         \item Developed an improved GAL function.
         \item Improves the low-performing outlier detection algorithms for diverse graphs.
     \end{itemize}       \vspace{1mm}      \end{minipage} 
     & 
     \begin{minipage}[t]{\linewidth}      \begin{itemize}[leftmargin=*]
         \item The algorithm indirectly increases the anomaly detection accuracy by including outliers rather for the regular values. 
     \end{itemize}       \vspace{1mm}      \end{minipage} \\
 
    %==========================================%    
     \hline
     Graph Neural Network-Based Anomaly Detection in Multivariate Time Series~\cite{deng2021graph} 
     &
     Developed a GDN to put together a structure learning technique with attention weights with GNN to explain anomalies in critical infrastructure systems. 
     & 
     \begin{minipage}[t]{\linewidth}      \begin{itemize}[leftmargin=*]
         \item Sensor embedding is computed to find deviations from normality.
         \item Graph attention mechanism is used for forecasting, unlike other approaches.
     \end{itemize}       \vspace{1mm}      \end{minipage}
     & 
     \begin{minipage}[t]{\linewidth}      \begin{itemize}[leftmargin=*]
         \item Sensor embedding learning plays a crucial role in the overall model learning, making it less robust.
     \end{itemize}       \vspace{1mm}      \end{minipage} \\
 
    %==========================================%
    \hline
    
    \end{tabular}
    }
    \caption{Summary of the existing literature on anomaly detection using GNN with scope, contribution, and limitations.}
    \label{table:anomaly_gnns}
}
\end{table*}

Chaudhary et al. ~\cite{chaudhary2019anomaly} developed a technique to detect point anomalies such as spam, fake reviews, or malicious activities within the social (Twitter ~\cite{twitter}) and email (Enron ~\cite{enron}) networks utilizing GNN. Findings conclude that `degree' and `between centrality' work best for capturing anomalous nodes.
% Li Paper - Relevance-Aware Anomalous Users Detection in Social Network via Graph Neural Network - 2021 [Done]
A similar social anomalous user (bots, spammers) detection approach was developed by Li et al.~\cite{li2021relevance}, where they used a Relevance-Aware Anomalous Users Detection model using GNN (RAU-GNN). A GCN-based relation fusion layer first generates an unfined multiple user-node relation graph by extracting from all users and relations. Then, a multi-head GAT-based embedding layer produces high-level embeddings for GNN to learn and identify anomalous users. Experimental high accuracy over Twitter~\cite{twitter} API and YelpChi ~\cite{yelpchi} dataset confirmed the claims.

% Ji Paper - An Anomaly Event Detection Method Based on GNN Algorithm for Multi-data Sources - June, 2021 [Done]
Ji et al.~\cite{ji2021anomaly} proposed a contextual anomalous event detection method considering dynamic multi-source data for secure critical infrastructure domain. Different feature-space social anomaly data were extracted and fused into a single feature space based on the abnormal score obtained through spectral clustering. A deep graph neural network (Deep-GNN) was trained over the fused data and, was tested using multi-source data from micro-blog~\cite{micro_blog} and Sina social platform~\cite{nr}. Test results compared with CNN and LSTM demonstrated robust and adaptive capabilities with an average accuracy of 95\%.
% Song paper - Spammer Detection Using Graph-level Classification Model of Graph Neural Network~\cite{song2021spammer} [Done]
Similarly, to prevent spamming and detect spammers in social media, Song et al.~\cite{song2021spammer} proposed a graph classification approach instead of node classification using GNN. The model is used to detect spammers from embedded interactive relationship graphs using a graph classification approach. Experiments were conducted using 10-fold cross-validation over the tagged.com dataset.

% Wu Paper - Graph Neural Networks for Anomaly Detection in Industrial Internet of Things - 2021 [Done]
% Wu et al.~\cite{wu2021graph} explained a few approaches using graph-level anomaly detection in the Industrial Internet of Things (IIoT) areas. The paper focused on point, contextual, and collective anomaly, established on three case studies, smart-transportation, smart-energy, and smart-factory industries. With regard to security, the contextual anomalies caused by electricity theft at power transformers were addressed using a GCN-based model on the Siamese network. Also, to check for cyberattack anomalies in smart factories, the authors created an n-layer GCN model with a SoftMax layer by converting the industrial communication network and the actions of IIoT devices and PLC into a graph. Experiments on the SGCC electricity theft dataset and SWaT, WADI, and CISS datasets for industrial cyberattacks justified the potential of the proposed approaches.

% Wang paper - One-class graph neural networks for anomaly detection in attributed networks - March, 2021 [Done]
Wang et al.\cite{wang2021one} proposed a One Class Graph Neural Network (OCGNN) for collective graph anomaly detection (GAD). OCGNN maps the nodes and relationships into a hypersphere to generate node embeddings close to the center. Hence, the common factors are identified by the GNN layer. GCN, GAT, or GraphSAGE can act as a GNN layer in this framework. Due to hypersphere learning, it can also be considered a natural extension of a one-class support vector machine (OCSVM). Cora~\cite{mccallum2000automating}, Citeseer~\cite{giles1998citeseer}, and Pubmed~\cite{sen2008collective} datasets were used for evaluation against twelve two-stage GAD methods and three SOTA GAE-based GAD approaches.
% Huang paper - One-class Temporal Graph Attention Neural Network for Dynamic Graph Anomaly Detection - 2021 [Done]
To extend the previous work for large dynamic graphs, Huang et al.\cite{huang2021one} proposed a one-class temporal graph attention neural network (OCTGAT) combining Temporal Graph Attention Network (TGAT)\cite{xu2020inductive} with OCGNN. On top of the previous approach, TGAT was used to extract the dynamic graph representations and aggregate structural and time-based attributes using Bocher's theorem with optimized computation. Anomalous nodes are detected by concatenating embeddings and feeding them to a feed-forward neural network. For such, the model requires to get trained with the non-anomalous dataset. According to the experiments over Wikipedia and Reddit datasets~\cite{kumar2019predicting}, OCTGAT surpassed the baseline approaches and can also execute edge classification and link prediction tasks.
% Shi Paper - Multi-layer Graph Neural Network-Based Random Anomalous Behavior Detection - 2021 [Done]
Shi et al.~\cite{shi2021multi} proposed another collective anomaly detection approach that facilitates data analysis by detecting random anomalous behavior. The model combines GCN and GAT to learn the abstract features with more attention paid to the differentiated influence of the local structure. A two-layer GCN learns connected node attributes and GAT identifies the influence of nodes at different positions from the parameterized probability distribution. Experiments on six real-world networks (OpenFlights~\cite{openflight}, Political blogs~\cite{adamic2005political}, Email-EU-core~\cite{email_eu_core}, Usairport, Jazz, and Highschool) alongside three benchmark models revealed its significant AUC edge over others.

% Protogerou Paper - A graph neural network method for distributed anomaly detection in IoT - 2020 [Done]
% To protect IoT devices, Protogerou et al.~\cite{protogerou2021graph} proposed a distributed multi-agent system (MAS) using GNN to exploit the collaborative and cooperative nature of intelligent agents for anomaly detection. The approach uses three-layered IoT infrastructure (IoT devices, Fog Nodes, and Cloud) to make an edge and node classifier that determines possible infection on a node and its corresponding edges based on their inter-relations. The approach leverages GNNs' distributed computing ability and minimizes resource consumption like bandwidth and energy. Generated datasets were used to conduct experiments based on real-time traffic distributions. SVM, Decision Tree, and Random Forest are three residual ML classifiers, along with two state-of-the-art approaches were used to test and compare against the proposed approach.

%Zhao Paper - GNN-based Graph Anomaly Detection with Graph Anomaly Loss - Aug, 2020 [Done]
In a parallel research direction, motivated by the low-performing random walk (RW) based training approaches for diverse graphs, Zhao et al.\cite{zhao2020gnn} proposed an improved graph anomaly loss (GAL) function that helps GNN to represent outlier and unexpected dense blocks with improved accuracy. GAL uses global group patterns identified by graph mining algorithms to evaluate similarity and adjust margins for minority classes. Graph outlier loss, bounded test error, and dense block loss are key attributes for such tasks. Experiments on BitCoin~\cite{kumar2016edge} and Weibo~\cite{jiang2016catching} datasets validated its improvement by around 10\%. GAL is also compatible with different GNN frameworks and has proven to keep the prediction error bounded by the proposed loss function. 
% Deng Paper - Graph Neural Network-Based Anomaly Detection in Multivariate Time Series - 2021 [Done]
On the other hand, to determine the root cause of the detected anomalies in high dimensional time series graphs, Deng et al.~\cite{deng2021graph} proposed a Graph Deviation Network (GDN) that utilizes structure learning with GNNs, alongside attention weights for critical infrastructure domain. The main components are sensor embedding, graph structure learning, graph attention-based forecasting, and graph deviation scoring. Experiments on real-world sensor dataset SWaT~\cite{mathur2016swat} \& WADI~\cite{ahmed2017wadi} justify its improved detection [F1-Score: 0.99 (SWaT) \& 0.98 (WADI)] and allow users to comprehend the root cause. We discuss persisting weaknesses and further research areas in Section \ref{subsection:anomaly_discussion}.
\vspace{-2mm}

\subsection{Vulnerability Detection} \label{subsection:vulnerability_summary}
    Software vulnerabilities are defects introduced in the source code during software development. These vulnerabilities left unattended can be exploited by an attacker to disrupt the client side including halting the crucial execution, integrating backdoors, remote file execution, and more. 
% With the relentless surge in vulnerability disclosures through Common Vulnerabilities and Exposures (CVE)~\cite{cve} and other platforms, the task of addressing them on time is a matter of utmost importance. 
During the pre-ML era, human experts were deployed to perform static, dynamic, or hybrid analysis on the source code to detect vulnerabilities~\cite{zheng2013path,sotirov2005automatic}. While these approaches were effective, they were mostly manual, which warranted intense labor and specialized expertise. With the advent of deep neural networks (DNNs), ML models could be trained on either expert-labeled patterns of vulnerability or fully automated techniques to detect vulnerability in source code. Vulnerability identification can be performed inter-procedural, which means identification of vulnerable code is analyzed on procedure or function level, whereas, intra-procedural vulnerability identification means the analysis is performed locally within a specific method scope. Several surveys summarize automated vulnerability detection using DNN techniques~\cite{shen2020survey, kushwaha2022systematic, hanif2021rise, rabheru2022hybrid}. Table~\ref{table:vulnerability_gnns} captures the focus of our survey to summarize recent developments using GNNs. 
 
\begin{table*}[!htbp]
\centering
{
    \footnotesize
    {
    \def\arraystretch{1.3}
    \begin{tabular}{ p{3cm}|p{3cm}|p{5cm}|p{3.5cm}  }
    \rowcolor{lightgray!20!}
    %==========================================%
    \hline
    \textbf{Paper Title} & 
    \textbf{Focus/Objective} & 
    \textbf{Contributions} & 
    \textbf{Limitations}\\
     %==========================================%
    \hline
    \rowcolor{cyan!20!}
    \multicolumn{4}{|c|}{\textbf{Research on Vulnerability Detection in Source Code}} \\
    %==========================================%

     %==========================================%
    \hline  
    Devign: Effective Vulnerability Identification by Learning Comprehensive Program Semantics via Graph Neural Networks~\cite{Zhou2019Devign}
    & 
    Source code vulnerability identification problem. 
    & 
    \begin{minipage}[t]{\linewidth}      \begin{itemize}[leftmargin=*]
         \item Novel composite code representation with AST, CFG, DFG, and NCS.
         \item Leveraged GGNN using 1-D CNN-based pooling (\textit{Conv} module) to perform graph-level classification task.
         % \item Attains accuracy 72.26\% and F1 score 73.26\% on LinuxKernel, QEMU, Wireshark, FFmpeg datasets.
      \end{itemize}       
      \vspace{1mm}     
      \end{minipage}
    &
     \begin{minipage}[t]{\linewidth}      \begin{itemize}[leftmargin=*]
        \item Complex Code representation leads to poor performance on large functions.
        \item Difficult to find programming language-specific parsers. %% This is mentioned in ReGVD
     \end{itemize}       \vspace{1mm}      \end{minipage} \\
    
     %==========================================%    
    \hline
    ReGVD: Revisiting Graph Neural Networks for Vulnerability Detection~\cite{Nguyen2022ReGVD}
    &
    Vulnerability identification as an inductive text classification problem.
    &
     \begin{minipage}[t]{\linewidth}      \begin{itemize}[leftmargin=*]
        \item Flat sequence of token representation allows PL-independent code graph.
        \item Introduced residual connection among GNN layers with sum and max poolings.
        % \item Attains accuracy 63.69\% on CodeXGLUE dataset.
     \end{itemize}       \vspace{1mm}      \end{minipage}
    &
     \begin{minipage}[t]{\linewidth}      \begin{itemize}[leftmargin=*]
        \item Code representation stage doesn't includes CFG or DFG, missing out on relevant structural information.
     \end{itemize}       \vspace{1mm}      \end{minipage} \\ 
     
     %==========================================%    
    \hline
    Combining Graph-Based Learning With Automated Data Collection for Code Vulnerability Detection~\cite{Wang2021combining}
    &
    Identifies software vulnerabilities at the function level from program source code.
    &
     \begin{minipage}[t]{\linewidth}      \begin{itemize}[leftmargin=*]
        \item Code representation using extended AST. 
        \item Leveraged GGNN with stacked GRU, which is passed to a standard fully-connected network to make a classification using a softmax layer. 
        % \item Model attains accuracy 92\% and F1 score 94\% on LinuxKernel, FFmpeg, ImageMagick, rdesktop, and OpenSC datasets.
     \end{itemize}       \vspace{1mm}      \end{minipage}
    &
    \begin{minipage}[t]{\linewidth}      \begin{itemize}[leftmargin=*]
        \item Pre-processing functions using extended AST adds significant complexity.
     \end{itemize}       \vspace{1mm}      \end{minipage} \\ 

     %==========================================%    
    \hline
    BGNN4VD: Constructing Bidirectional Graph Neural-Network for Vulnerability Detection~\cite{cao2021bgnn4vd}
    &
    Vulnerability detection approach by constructing a Bidirectional Graph Neural-Network (BGNN)
    &
     \begin{minipage}[t]{\linewidth}      \begin{itemize}[leftmargin=*]
        \item Proposed CCG to capture source-code syntax and semantics.
        \item Used BGNN to process information and classify using a CNN.
        % \item Model attains accuracy 74.7\% and F1 score 76.8\% on LinuxKernel, FFmpeg, Wireshark, and Libav datasets. 
     \end{itemize}       \vspace{1mm}      \end{minipage}
    &
     \begin{minipage}[t]{\linewidth}      \begin{itemize}[leftmargin=*]
        \item Accuracy is specific to reported vulnerabilities.
        \item Non-reliable for large datasets.
     \end{itemize}       \vspace{1mm}      \end{minipage} \\ 
     
     %==========================================%    
    \hline
    DeepWukong: Statically Detecting Software Vulnerabilities Using Deep Graph Neural Network~\cite{Cheng2021DeepWukong}
    &
    Precise interprocedural analysis for vulnerability detection in real-world programs. 
    &
     \begin{minipage}[t]{\linewidth}      \begin{itemize}[leftmargin=*]
        \item Code representation is performed by generating PDG containing control and data dependency of the program.
        \item Leveraged GCN, GAN, and k-dimensional GNNs with top-k pooling.
        % \item Model attains accuracy score 97.4\% and F1 score 95.6\% on SARD, lua, redis datasets.
     \end{itemize}       \vspace{1mm}      \end{minipage}
    &
    \begin{minipage}[t]{\linewidth}      \begin{itemize}[leftmargin=*]
      \item Datasets are labeled by domain experts, possibility of missing corner cases.
      \item Limited to the top 10 vulnerabilities in C/C++ programs.
    \end{itemize}       \vspace{1mm}      \end{minipage} \\
  
    %==========================================%
    \hline
    CSGVD: A deep learning approach combining sequence and graph embedding for source code vulnerability detection~\cite{tang2023csgvd}
    &
    Function-level vulnerability detection as a graph binary
    classification task.&
     \begin{minipage}[t]{\linewidth}      \begin{itemize}[leftmargin=*]
        \item Code representation using CFG.
        \item Leveraged PE-BL module, a GNN layers with residual connectivity, and a graph-level mean biaffine attention (M-BFA) pooling to learn graph representation.
     \end{itemize}       \vspace{1mm}      \end{minipage}
    &
     \begin{minipage}[t]{\linewidth}      \begin{itemize}[leftmargin=*]
        \item Information from DFG and AST was not considered.
        \item Lack of interprocedural information to understand the nature of vulnerability.
     \end{itemize}       \vspace{1mm}      \end{minipage} \\

      %==========================================%    
    \hline
    A hybrid graph neural network approach for detecting PHP vulnerabilities~\cite{rabheru2022hybrid}
    &
    Vulnerability detection by capturing contextual information to reduce false positives and false negatives.
    &
     \begin{minipage}[t]{\linewidth}      \begin{itemize}[leftmargin=*]
        \item Used intraprocedural CFGs to capture semantic code dependencies.
        \item Leveraged bidirectional GRUs, GCN, and edge pooling.
        % \item The model attains accuracy 96.56\% and F1 score 96.11\% on SARD \& Real-World Projects File Level. 
     \end{itemize}       \vspace{1mm}      \end{minipage}
    &
     \begin{minipage}[t]{\linewidth}      \begin{itemize}[leftmargin=*]
        \item The model works on only PHP source codes.
        \item Experiments are run on simplistic SARD  dataset.
     \end{itemize}       \vspace{1mm}      \end{minipage} \\

     %==========================================%    
    \hline 
    Predicting Vulnerability Inducing Function Versions Using Node Embeddings and Graph Neural Networks~\cite{sefa2022}
    &
    Vulnerability prediction model, that runs after every code change and identifies vulnerability-inducing functions. 
    &
     \begin{minipage}[t]{\linewidth}      \begin{itemize}[leftmargin=*]
        \item Code representation using AST
        \item Leveraged GraphSAGE, GCN with average pooling
        and max pooling. 
        % \item F1 score 74.4 on Wireshark dataset.
     \end{itemize}       \vspace{1mm}      \end{minipage}
    &
    \begin{minipage}[t]{\linewidth}      \begin{itemize}[leftmargin=*]
        \item Tested on only one dataset.
     \end{itemize}       \vspace{1mm}      \end{minipage} \\

      %==========================================%    
    \hline
    \rowcolor{cyan!20!}
    \multicolumn{4}{c}{\textbf{Research on Vulnerability Detection in Smart Contracts}} \\
     %==========================================%

     %==========================================%
    \hline
    Smart Contract Vulnerability Detection Using
    Graph Neural Networks~\cite{zhuang2020smart} 
    &
    Fully automated vulnerability analyzer for smart contracts.
    & 
     \begin{minipage}[t]{\linewidth}      \begin{itemize}[leftmargin=*]
        \item Code represented using contract graph with the consideration of temporal ordering. Node types include, i.e., major nodes, secondary nodes, and fallback nodes.
        \item Leveraged DR-GCN and novel temporal message propagation network TMP. 
    
     \end{itemize}       \vspace{1mm}      \end{minipage} &
     \begin{minipage}[t]{\linewidth}      \begin{itemize}[leftmargin=*]
        \item Source code is considered as a text sequence instead
        of semantic blocks, failing to highlight critical variables in
        the data flow.
     \end{itemize}       \vspace{1mm}      \end{minipage} \\
     
     %==========================================%    
    \hline
    
    \end{tabular}
    }
    \caption{
    Summary of the existing literature on vulnerability detection using GNN with scope, contribution, and limitations.}
    \label{table:vulnerability_gnns}
}
\end{table*}

To automate the process of vulnerability detection, one major decision is to learn comprehensive program semantics which can capture both data and logical flow of the source code. Zhou et al.~\cite{Zhou2019Devign} proposed Devign that solves vulnerability identification problems as graph-level classification where the model learns from composite programming representation of the source code, such as (1) abstract syntax tree (AST), (2) control flow graph (CFG), (3) data flow graph (DFG), and (4) natural code sequence (NCS). The authors proposed a novel 1-D CNN-based pooling mechanism to translate node and link level classification for end-to-end graph level classification. Evaluation over LinuxKernel, QEMU, Wireshark, and FFmpeg datasets~\cite{devign_dataset} attains accuracy of 72.26\% and F1 score of 73.26\%. The results outperform state-of-the-art~\cite {Du2019leopard} by a higher average of 10.51\% accuracy and 8.68\% F1 score. Devign is considered in the literature as one of the first and strongest baseline models for inter-procedural vulnerability detection. 
Nguyen et al.~\cite{Nguyen2022ReGVD} claim the use of Devign~\cite{Zhou2019Devign} is impractical for complex programming languages, source codes, and libraries due to the difficult pre-processing step to transform source code into a multi-edged graph. The authors proposed ReGVD, a language-independent vulnerability identification mechanism, that represents the graph as a flat sequence of tokens. The represented graph is initialized by embeddings from pre-trained programming language (PL) models like CodeBERT or GraphCodeBERT. The GNN node-level embeddings are updated in each iteration by recursive aggregation using a residual connection. Finally, a graph-level readout layer aggregates node embeddings globally for each graph. Evaluation results over CodeXGLUE~\cite{lu2021codexglue} show ReGVD achieving accuracy improvement of 1.61\% over Devign. 
To unleash the full potential of DL-based vulnerability detection, Wang et al~\cite{Wang2021combining} proposed FUNDED, a mechanism that encodes multiple code relationships into different relation graphs, to learn relation-specific functions and their associated vulnerability. The model leverages GGNN with stacked GRU and a fully connected network to make classification. Evaluation results over LinuxKernel, FFmpeg, ImageMagick, rdesktop, and OpenSC datasets~\cite{funded} show an improvement of 12.4\% in accuracy over Devign. 
Tang et al~\cite{tang2023csgvd} proposed CSGVD that uses a PE-BL module which inherits and enhances the knowledge from the pre-trained language model for node embedding and a BiLSTM to aggregate the local semantic information within a node. Additionally, a mean bi-affine attention pooling (M-BFA) for graph embedding. The evaluation results over the CodeXGLUE~\cite{lu2021codexglue} dataset were compared with multiple baselines proving an improvement of 4.41\% accuracy over Devign. 
Cao et al~\cite{cao2021bgnn4vd} proposed another bidirectional GNN to perform vulnerability identification using a code composite graph (CCG) by combining the AST, CFG, and DFG that captures various source code syntax and semantic information. Evaluation over LinuxKernel, FFmpeg, Wireshark, and Libav datasets~\cite{bgnn4vd} reported an accuracy of 74.7\% and F1-score of 76.8\%.   

Cheng et al.~\cite{Cheng2021DeepWukong} introduced an interprocedural static vulnerability identification tool tailored for C/C++. This tool calculates control and data dependencies over CFG and Value-Flow Graphs (VFG), considering pointer aliases information. Additionally, it constructs the Program Dependence Graph (PDG). The model achieved an accuracy score of 97.4\% and an F1-score of 95.6\% across multiple datasets, including SARD~\cite{sard}, lua~\cite{lua}, and redis~\cite{redis}. In another context, Rabheru et al.~\cite{rabheru2022hybrid} tackled intra-procedural vulnerability detection in PHP. They employed a token-based approach, parsing linear token sequences to capture syntactic dependencies in source code using CFGs. Their model delivered an accuracy of 96.56\% and an F1-score of 96.11\% on both the SARD~\cite{sard} dataset and real-world projects. These results were achieved by harnessing bidirectional Gated Recurrent Units (GRUs) in conjunction with Graph Convolutional Networks (GCNs). Şahin et al.~\cite{sefa2022} proposed a method for detecting change-level vulnerable code using Abstract Syntax Trees (AST) and incorporating GCNs. The model achieved an F1-score of 74.4\% when evaluated on the Wireshark dataset~\cite{wireshark-dataset}.

Another avenue of vulnerability detection includes blockchain systems and other cryptocurrency networks. The rise of online wallets has popularized decentralized online ledgers for monitoring financial transactions. To ensure trustworthy and credible online transactions, organizations encode rules and policies in the form of source code, commonly referred to as \textit{smart contracts}. Similar to any code snippet, smart contracts can be susceptible to vulnerabilities~\cite{Cai2023combine}. Zhuang et al.~\cite{zhuang2020smart} presented a novel approach involving a degree-free Graph Convolutional Network (GCN) and a Temporal Message Propagation network (TMP) for detecting vulnerabilities in smart contracts. Their model achieved remarkable accuracy, scoring 84.48\%, 83.45\%, and 74.61\% for identifying reentrancy, timestamp dependence, and infinite loop vulnerabilities. In the context of our survey, we refrain from an in-depth exploration of the literature on smart contract vulnerability detection due to the existance of substantial body of work, including comprehensive surveys~\cite{qian2022smart, chu2023survey, zhang2021recent} that offer profound insights into the landscape of smart contract vulnerabilities and the application of GNNs for their detection. In Section \ref{subsection:vulnerability_discussion}, we delve further into the persisting loopholes and areas for improvement. 

\subsection{Intrusion Detection} \label{subsection:intrusion_summary}
    % -- Basic Into about Intrusion detection -- 
% NIDS \& NIPS
% https://www.sciencedirect.com/topics/computer-science/network-based-intrusion-detection-system#:~:text=A%20network%2Dbased%20intrusion%20detection,the%20traffic%20they%20monitor%3B%20Fig.

% Network Intrusion Detection Systems (NIDS) are mainly of two types, signature-based and intrusion detection-based systems.

Intrusion detection refers to the process of monitoring and analyzing computer systems or networks to identify unauthorized or malicious activities. It involves the use of various TTPs to detect and respond to potential security breaches, such as unauthorized access, malware infections, or end-point security. It can be considered as a countermeasure against the installation phase of the CKC where the attacker tries to establish a backdoor for persistent communication. Intrusion Detection Systems (IDS) are mainly of two types, Network-based (NIDS) and Host-based (HIDS). NIDS monitors network traffic for suspicious patterns or anomalies, while HIDS focuses on individual systems or endpoints. There have been numerous studies on IDS utilizing GNN capabilities and is still an active area of research~\cite{bilot2023graph}. Some notable ones are discussed below with a summarized Table~\ref{table:intrusions_gnns}. 

 \begin{table*}[!htbp]
\centering
{
    \footnotesize
    {
    \def\arraystretch{1.3}
    \begin{tabular}{ p{3cm}|p{3cm}|p{5cm}|p{3.5cm}  }
    \rowcolor{lightgray!20!}
    %==========================================%
    \hline
    \textbf{Paper Title} & 
    \textbf{Focus/Objective} & 
    \textbf{Contributions} & 
    \textbf{Limitations}\\
    %==========================================%
    \hline
    \rowcolor{cyan!20!}
    \multicolumn{4}{|c|}{\textbf{Network Intrusion Detection System (NIDS)}} \\
    %==========================================%

     %==========================================%
     \hline
     E-GraphSAGE: A Graph Neural Network based Intrusion Detection System for IoT~\cite{lo2022graphsage}
     & 
     Developed GNN-based intrusion detection system for IoT domain by capturing edge features and topological information.
     & 
     \begin{minipage}[t]{\linewidth}      \begin{itemize}[leftmargin=*]
         \item Considers flow-based features and topological patterns to account for interconnected patterns of the network flow.
         \item Modified GraphSAGE algorithm with mini-batch allows the model to consider edges with increased efficiency.
     \end{itemize}       \vspace{1mm}      \end{minipage}
     & 
     \begin{minipage}[t]{\linewidth}      \begin{itemize}[leftmargin=*]
         \item Neighborhood sampling can be used to improve the run-time of the model.
         \item Investigation on explainable GNN algorithms is needed to get model output insights.
     \end{itemize}       \vspace{1mm}      \end{minipage}\\

      %==========================================%
     \hline
     Graph-based Solutions with Residuals for Intrusion Detection: the Modified E-GraphSAGE and E-ResGAT Algorithms~\cite{chang2021graph}
     &
     Integrated EGraphSAGE and E-ResGAT to incorporate residual connections and attention mechanism to consider class imbalance.
     & 
     \begin{minipage}[t]{\linewidth}      \begin{itemize}[leftmargin=*]
         \item Incorporated residual learning with EGraphSAGE to enhance performance.
         \item Applied attention mechanism of GAT while considering residual learning and edge features to improve efficiency further.
     \end{itemize}       \vspace{1mm}      \end{minipage}
     &
     \begin{minipage}[t]{\linewidth}      \begin{itemize}[leftmargin=*]
         \item Optimizing the construction of the batch neighborhood process can speed up the model performance.
         \item More imbalanced data sets need to be tested.
     \end{itemize}       \vspace{1mm}      \end{minipage}\\

     %==========================================%
     \hline
     Unveiling the potential of Graph Neural Networks for robust Intrusion Detection~\cite{pujol2022unveiling}
     & 
     Developed a GNN NIDS model that can process and learn from structural relationships and flow patterns to identify attacks.
     &
     \begin{minipage}[t]{\linewidth}      \begin{itemize}[leftmargin=*]
         \item Proposed host-connection graphs, to capture structural flow relationships.
         \item Proposed a GNN model that uses a non-standard message-passing architecture to learn from host-connection graphs.
     \end{itemize}       \vspace{1mm}      \end{minipage}
     &
     \begin{minipage}[t]{\linewidth}      \begin{itemize}[leftmargin=*]
         \item Was only tested on CIC-IDS2017 dataset and synthesized adversarial attack scenarios.
     \end{itemize}       \vspace{1mm}      \end{minipage}\\

      %==========================================%
      \hline
      \rowcolor{cyan!20!}
      \multicolumn{4}{|c|}{\textbf{Host Intrusion Detection System (HIDS)}} \\
      %==========================================%  

     %==========================================%
     \hline
     Using Graph Representation in Host-Based Intrusion Detection~\cite{hu2021using}
     & 
     Graph representation learning-based host intrusion detection system (HIDS).
     &
     \begin{minipage}[t]{\linewidth}      \begin{itemize}[leftmargin=*]
         \item Used graph representation learning to develop HIDS.
         \item Developed a sequence embedding method using graph structures to model a finite number of sequence items and represents the structural relationships between them.
     \end{itemize}       \vspace{1mm}      \end{minipage}
     & 
     \begin{minipage}[t]{\linewidth}      \begin{itemize}[leftmargin=*]
         \item Cannot generate graphs for sequences with multiple attributes or predefined rules.
         \item Performance improvement of GRSE for larger full graphs needs to be explored. 
     \end{itemize}       \vspace{1mm}      \end{minipage}\\

     %==========================================%
     \hline
     THREATRACE: Detecting and Tracing Host-Based Threats in Node Level Through Provenance Graph Learning~\cite{wang2022threatrace}
     & 
     An anomaly-based HIDS at the system entity level without prior knowledge of any attack pattern.
     & 
     \begin{minipage}[t]{\linewidth}      \begin{itemize}[leftmargin=*]
         \item Used GraphSAGE to learn benign entities' role in provenance graph.
         \item The model works in a real-time manner and can scale for long-term hosts.
         
     \end{itemize}       \vspace{1mm}      \end{minipage}
     &
     \begin{minipage}[t]{\linewidth}      \begin{itemize}[leftmargin=*]
         \item Require benign data for training and cannot guarantee robustness.
         \item May fail against poisoning and graph backdoor attacks.
     \end{itemize}       \vspace{1mm}      \end{minipage}\\

    %==========================================%
    \hline
    \end{tabular}
    }
    \caption{
    Summary of the existing literature on intrusion detection on GNN with scope, contribution, and limitations.}
    \label{table:intrusions_gnns}
}
\vspace{-5mm}
\end{table*}

% Lo E-GraphSAGE Paper - E-GraphSAGE: A Graph Neural Network based Intrusion Detection System [Done]
Weng et al.~\cite{lo2022graphsage} presented NIDS using GNN to classify malicious network flows. The model captures edge and topological information in IoT networks for classification. The authors enhanced the GraphSAGE~\cite{hamilton2017inductive} algorithm to directly exploit the structural information of the network flow and encode it in a graph. The main changes in E-GraphSAGE are in the algorithm input, the message passing/aggregator function, and the output where the network flow edges have been considered instead of nodes. Experiments over the datasets created in~\cite{koroniotis2019towards, alsaedi2020ton_iot, sarhan2020netflow} achieved an F-1 score of 0.97 for binary and 0.87 for multi-class classification.
% Chang Modified E-GraphSAGE Paper - Graph-based Solutions with Residuals for Intrusion Detection: the Modified E-GraphSAGE and E-ResGAT Algorithms [Done]
Chang et al.~\cite{chang2021graph} further enhanced E-GraphSAGE to integrate residual learning and attention mechanism to increase efficiency. The authors utilized E-GraphSAGE with residual learning to target minority class imbalance and an edge-based residual graph attention network (E-ResGAT) to improve efficiency. They also proposed a fixed neighborhood in edge sampling and selection for aggregation using attention mechanism. Evaluation over CIC-DarkNet~\cite{cic_daknet}, ToN-IoT~\cite{ton-iot}, UNSW-NB15~\cite{unsw_nb15}, and CSE-CIC-IDS ~\cite{cse_cic_ids, sharafaldin2018toward} datasets proved an overall increase in accuracy.
% Pujol Paper - Unveiling the potential of Graph Neural Networks for robust Intrusion Detection [Done]
To make the model robust against changing networks and evolving adversarial attacks, Pujol-Perich et al.~\cite{pujol2022unveiling} focused on the structural patterns of the attack by analyzing the flow features independently with intra-relations. They proposed a host-connection (HC) graph to keep the flow records and capture meaningful information with a GNN model based on a non-standard message-passing architecture capable of learning from the HC graph to recognize the structural flow patterns of attacks. Evaluation over CIC-IDS2017~\cite{sharafaldin2018toward} dataset against two adversarial attacks that modify packet size and inter-arrival times justified its superiority. While other traditional ML models degrade their accuracy (F1) up to 50\%, the proposed model maintained its accuracy throughout.

% Hu paper - Using Graph Representation in Host-Based Intrusion Detection~\cite{hu2021using}
For HIDS from the system calls, Hu et al.~\cite{hu2021using} proposed a graph-based model that works by constructing graphs from system call traces for graph classification to detect intrusion. Graph random state embedding (GRSE) is another novel approach developed by the authors to generate graph embeddings, including graph transform, random state walk, subgraph extraction, and graph pooling. Hence, classification is done based on the embedded graph topology. Experiments on the ADFA-LD~\cite{creech2013semantic} dataset justify the performance improvement with the developed embedding.
% Wang paper - THREATRACE: Detecting and Tracing Host-Based Threats in Node Level Through Provenance Graph Learning~\cite{wang2022threatrace} [Done]
Wang et al.~\cite{wang2022threatrace} developed another real-time HIDS framework based on system provenance graphs. The model utilized GraphSAGE algorithms for aggregation and node embedding generation, where each system entity is considered as nodes, and the system calls as edges. The model requires only benign data for training and works in the anomaly identification principle. Experiments over five public datasets (StreamSpot~\cite{streamspot}, Unicorn SC-1 and SC-2~\cite{unicorn}, DARPA TC \#3 and \#5~\cite{darpa}) against seven state-of-the-art HIDS models proved its potential. Persisting loopholes and improvement areas are discussed in Section~\ref{subsection:intrusion_discussion}.
\vspace{-3mm}

\subsection{Malware Detection} \label{subsection:malware_summary}
    Malware detection is a crucial countermeasure against the Command and Control (C2) phase. It phase involves attacker utilizing the persistent communication channels, set up by installed malware in the compromised systems to execute remote instruction. Hence, anti-malware solutions employ diverse methods to detect and block installed malware activities by learning system behavioral patterns through ML models. To capture behavioral relations, GNN has proven to be an attractive choice that enables the identification of malware presence by addressing suspicious behaviors. Additionally, behavior analysis can also indicate the presence of zero-day malwares.
\begin{table*}[!htbp]
\centering
{
    \footnotesize
    {
    \def\arraystretch{1.3}
    \begin{tabular}{ p{3cm}|p{3cm}|p{5cm}|p{3.5cm}  }
    \rowcolor{lightgray!20!}
    %==========================================%
    \hline
    \textbf{Paper Title} & 
    \textbf{Focus/Objective} & 
    \textbf{Contributions} & 
    \textbf{Limitations}\\
    %==========================================%
    \hline
    \rowcolor{cyan!20!}
    \multicolumn{4}{|c|}{\textbf{General Malwares}} \\
    %==========================================%

     %==========================================%
     \hline
     Heterogeneous Graph Matching Networks: Application to Unknown Malware Detection~\cite{wang2019heterogeneous}
     & Developed a heterogeneous graph-matching network model to learn the graph representation and similarity metric simultaneously based on program execution.
     &
     \begin{minipage}[t]{\linewidth}      \begin{itemize}[leftmargin=*]
         \item Developed an invariant graph modeling to capture heterogeneous interactions.
         \item HAGNE was used to learn from heterogeneous invariant graphs. 
         \item Siamese network was used to perform similarity scoring.
     \end{itemize}       \vspace{1mm}      \end{minipage}  
     &
     \begin{minipage}[t]{\linewidth}      \begin{itemize}[leftmargin=*]
         \item Accuracy needs to be improved further.
         \item Evaluation with more robust attack scenarios is needed for practical deployment.
     \end{itemize}       \vspace{1mm}      \end{minipage}  \\
    
     % \hline
     % Automating Botnet Detection with Graph Neural Networks~\cite{zhou2020automating} 
     % & 
     % Developed GNN model to identify botnets from massive background internet communication graphs in an automated manner.
     % & 
     % \begin{minipage}[t]{\linewidth}      \begin{itemize}[leftmargin=*]
     %     \item Considered on detection of P2P botnets.
     %     \item The model solely learns from graph topology and does not require attributed graphs.
     %     \item Generated synthesized botnet training data with underlying communication patterns overlaid on large-scale real networks.
     % \end{itemize}       \vspace{1mm}      \end{minipage}  
     % & 
     % \begin{minipage}[t]{\linewidth}      \begin{itemize}[leftmargin=*]
     %     \item Only limited to botnets and does not cover other similar attack scenarios like DDoS, prefix hijacking, etc.
     % \end{itemize}       \vspace{1mm}      \end{minipage}  \\
    
     \hline
     Classifying malware represented as control flow graphs using deep graph convolutional neural network~\cite{yan2019classifying} 
     & Detect malware from CFG using DGCNN to be deployable in a variety of operational environments.
     & 
     \begin{minipage}[t]{\linewidth}      \begin{itemize}[leftmargin=*]
         \item Used DGCNN for CFG analysis and classification.
         \item Developed generically to be deployable in the cloud and can be used by a common user.
     \end{itemize}       \vspace{1mm}      \end{minipage} 
     &
     \begin{minipage}[t]{\linewidth}      \begin{itemize}[leftmargin=*]
         \item Requires a high training time.
         \item Requires testing with the latest malware samples for robustness.
     \end{itemize}       \vspace{1mm}      \end{minipage} \\
    
     \hline
     Classifying Packed Malware Represented as Control Flow Graphs using Deep Graph Convolutional Neural Network~\cite{hua2020classifying} 
     &
     Develop malware classifier using DGCNN while considering the unpacked and local CFG of applications.
     & 
     \begin{minipage}[t]{\linewidth}      \begin{itemize}[leftmargin=*]
         \item Developed algorithm to strip from packed CFG to obtain unpacked local CFG.
         \item Used DGCNN to learn and classify malware from unpacked block CFGs. 
     \end{itemize}       \vspace{1mm}      \end{minipage} 
     & 
     \begin{minipage}[t]{\linewidth}      \begin{itemize}[leftmargin=*]
         \item Different adversarial CFG characteristics are required to be tested with the approach for robust applicability.
     \end{itemize}       \vspace{1mm}      \end{minipage} \\
    
     \hline
     Intelligent malware detection based on graph convolutional network~\cite{li2022intelligent} 
     &
     Develop a malware classifier using GCN to adapt to the different malware characteristics.
     & 
     \begin{minipage}[t]{\linewidth}      \begin{itemize}[leftmargin=*]
         \item Used directed cyclic graph, constructed from API call sequence. 
         \item Used Markov chain and PCA for feature analysis.
         \item Used GCN for malware classification.
     \end{itemize}       \vspace{1mm}      \end{minipage} 
     & 
     \begin{minipage}[t]{\linewidth}      \begin{itemize}[leftmargin=*]
         \item Further research on GCN is required to make the model more adaptive with the development of novel malware.
         \item Cost optimization also needs to be addressed.
     \end{itemize}       \vspace{1mm}      \end{minipage} \\

    %==========================================%
    \hline
    \rowcolor{cyan!20!}
    \multicolumn{4}{|c|}{\textbf{Android Malwares}} \\
    %==========================================%
    
     \hline
     Android Malware Detection via Graph Representation Learning~\cite{feng2021android} 
     &
     Malware detection using graph representation learning, combined with NLP techniques.
     &
     \begin{minipage}[t]{\linewidth}      \begin{itemize}[leftmargin=*]
         \item Lightweight static analysis using graph representation learning.
         \item Word2Vec generates function embedding to represent code characteristics.
         \item Captures semantic information from call graphs without expert knowledge.
     \end{itemize}       \vspace{1mm}      \end{minipage} 
     &
     \begin{minipage}[t]{\linewidth}      \begin{itemize}[leftmargin=*]
         \item Cannot extract intra-function Smali instructions against obfuscation.
         \item Cannot create precise approximate call graphs for large-scale malware.
     \end{itemize}       \vspace{1mm}      \end{minipage} \\

    %==========================================%
    \hline
    NF-GNN: Network Flow Graph Neural Networks for Malware Detection and Classification~\cite{busch2021nf} 
     & 
     Developed a network flow feature-based GNN model for malware detection and classification.
     & 
     \begin{minipage}[t]{\linewidth}      \begin{itemize}[leftmargin=*]
         \item Extracted directed edge attribute flow graphs from a set of network flows.
         \item Developed a GNN model to learn from graph topology and edge attributes for malware classification.
     \end{itemize}       \vspace{1mm}      \end{minipage} 
     & 
     \begin{minipage}[t]{\linewidth}      \begin{itemize}[leftmargin=*]
         \item Additional network architectures such as attention, model temporal dynamics, and explainability need to be considered.
     \end{itemize}       \vspace{1mm}      \end{minipage} \\

     %==========================================%
     \hline
     
     Graph Neural Network-based Android Malware Classification with Jumping Knowledge~\cite{weng2022graph} 
     & 
     Android malware detection through FCGs using GNN-JK.
     & 
     \begin{minipage}[t]{\linewidth}      \begin{itemize}[leftmargin=*]
         \item Used GNN with JK technique to address GNN over-smoothing problem. 
         \item Topological information embedded in FCGs was utilized for malware classification.
     \end{itemize}       \vspace{1mm}      \end{minipage} 
     & 
     \begin{minipage}[t]{\linewidth}      \begin{itemize}[leftmargin=*]
         \item The developed model is not optimized at all and exploring other GNN architectures might help.
     \end{itemize}       \vspace{1mm}      \end{minipage} \\

     %==========================================%
     \hline
     \rowcolor{cyan!20!}
     \multicolumn{4}{|c|}{\textbf{Industrial \& IoT Malwares}}\\
     %==========================================%

     %==========================================%
     \hline
     Cross-Architecture Internet-of-Things Malware Detection Based on Graph Neural Network~\cite{li2021cross} 
     & 
     Developed a cross-architecture IoT malware detection method using GNN.
     & 
     \begin{minipage}[t]{\linewidth}      \begin{itemize}[leftmargin=*]
         \item Used NLP to extract semantic information and generate function embedding.
         \item Used GraphSAGE to learn the structural information by GNN.
         \item Used MLP with softmax to classify malicious binary files.
     \end{itemize}       \vspace{1mm}      \end{minipage} 
     &
     \begin{minipage}[t]{\linewidth}      \begin{itemize}[leftmargin=*]
         \item Pre-processing phase using IDA pro is time-consuming and requires significant acceleration.
         \item Need to consider more features to increase robustness.
     \end{itemize}       \vspace{1mm}      \end{minipage} \\
     
    \hline
    \end{tabular}
    }
    \caption{
    Summary of the existing literature on malware detection using GNN with scope, contribution, and limitations.}
    \label{table:malware_gnns}
}
\end{table*}

Wang et al.~\cite{wang2019heterogeneous} developed a behavior-based malware detection approach (MatchGNet) using a heterogeneous graph matching mechanism. Invariant graph modeling (IGM) captures heterogeneous system interactions, then a hierarchical attention graph neural encoder (HAGNE) is employed to learn the representations. Finally, a similarity learning (SL) model via Siamese Network~\cite{zagoruyko2015learning} trains the parameters and determines the similarity between the unknown and a benign program. Experiments revealed an increased accuracy of up to 97\% while outperforming other state-of-the-art baseline models (LR, SVM, MLP, GCN, GraphSAGE) with 50\% less false-positive (FP) rates at lower training time and cost.
% Zhou Paper - Automating botnet detection with graph neural networks - 2020 [Done]
% To address P2P botnet nodes, Zhou et al.~\cite{zhou2020automating} developed a GNN model that classifies botnets automatically by their topological characteristics within a massive internet communication graph. For the experiment, the authors first generated synthesized botnet traffic over the CAIDA dataset with four synthetic (DE BRUIJN, KADEMLIA, CHORD, and LEFT-CHORD) and two real-world (C2 and P2P) botnet topologies replicating diverse underlying communication patterns. Comparison against BotGrep and logistic regression (LR) proved its superiority.
% Yan et al.\cite{yan2019classifying} - Classifying malware represented as control flow graphs using deep graph convolutional neural network - 2019 [Done]
Yan et al.\cite{yan2019classifying} proposed a malware classification tool that utilizes the graph mining capabilities of Deep Graph Convolution Network (DGCNN). Since the CFG has a heterogeneous data structure, it is represented as a tensor of variable size, requiring a graph machine learning approach. Experiments on MSKCFG~\cite{ronen2018microsoft} and YANCFG~\cite{yancfg, virustotal} datasets were conducted and evaluated with 5-fold cross-validation. According to the evaluation, the model achieved an average F-1 score of 0.97 in MSKCFG and around 0.8 in YANCFG dataset. Due to its generic approach, it can also be deployed in the cloud for real-time classification.
% Hua paper - Classifying Packed Malware Represented as Control Flow Graphs using Deep Graph Convolutional Neural Network - 2020 [Done]
To avoid detection during execution, another technique malware uses is packing, which generates different CFGs. Hua et al.~\cite{hua2020classifying} proposed to strip the unpacked CFG into local CFG for final classification using DGCNN. The unpack function calls do not relate to any malware local functions and vice-versa. Hence, the unpacked CFG can be stripped to the local CFG for classification. Finally, DGCNN learns the malicious local CFG for classification. Experiments covering six malware families from VirusShare~\cite{virusshare} with 10-fold cross-validation yield an overall accuracy of 96.4\%.
% Li paper - Intelligent malware detection based on graph convolutional network~\cite{li2022intelligent}  [Done]
To address the evolving malware characteristics, Li et al.~\cite{li2022intelligent} proposed a malware classifier using GCN. The objective is to learn different characteristics from directed API call graphs. The notable aspect of this approach is use of Markov chain and PCA for graph feature extraction. Experiments over VirusTotal~\cite{virustotal} and VirusShare~\cite{virusshare} concluded the highest accuracy of 98.32\% with less FPR than other state-of-the-art models.

% Feng Paper - Android Malware Detection via Graph Representation Learning - Feb<>June - 2021 [Done]
To detect Android malware, Feng et al.~\cite{feng2021android} proposed a lightweight static analysis approach (CGDroid) that analyzes the source code to extract high-level semantic information. It constructs an approximate call graph from function invocation relationships and then pulls out intra-function attributes like permissions, security level, Smali instructions, etc. Word2Vec-embedded call graphs allow unsupervised training and classification using GNN. Evaluation over Drebin~\cite{arp2014drebin} and AndroZoo~\cite{allix2016androzoo} datasets suggests promising capabilities with an accuracy of 97.1\% for multi-class (20) classification.
% Busch Paper - NF-GNN: Network Flow Graph Neural Networks for Malware Detection and Classification - June 2021 [Done]
To leverage rich communication patterns and dynamically detect Android malware using network flow graphs while apprehending endpoint pairs using GNN, Busch et al.~\cite{busch2021nf} developed three derived models: graph classifier, graph autoencoder, and one-class graph neural network (OCGNN). It extracts a communication graph for each execution of a candidate application and then classifies them using edge features. Evaluation results over CICAndMal2017~\cite{lashkari2018toward} dataset improved recall by 4.12\%, 14.41\%, and 24.78\% for binary, category, and family classification while obtaining 95\% detection accuracy. Additionally, the ablation study helped evaluate the influence of feature sets and network layers. 
% Lo Paper - Graph Neural Network-based Android Malware Classification with Jumping Knowledge - June 2022 [Done]
Lo et al.~\cite{weng2022graph} proposed another network-based approach using GNN with jumping knowledge (JK) by utilizing Android function call graphs (FCGs) to obtain meaningful intra-procedural call patterns. GCN, GraphSAGE, and Graph Isomorphism Network (GIN)~\cite{xu2019how} built a three-phase classification approach. Apart from traditional feature extraction and classification, JK is used in training concatenation layer to obtain graph embedding while preventing over-smoothing. Evaluations over 24 malware families from Malnet-Tiny~\cite{freitas2020large} and Drebin~\cite{arp2014drebin} datasets proved to boost accuracy by 8\%.

% Li paper - Cross-Architecture Internet-of-Things Malware Detection Based on Graph Neural Network - 2021 [Done]
An architecture-independent malware detection approach for IoT devices focusing on FCG was proposed by Li et al.~\cite{li2021cross}. The model can consider structural semantics information independent of the underlying architecture. Experiments covering five different processor architectures (MIPS, ARM, PowerPC, X86\_64, i386) with datasets~\cite{virusshare} covering diverse malware families over the years were used for evaluation and achieved an accuracy of 99.61\%. 
Apart from solo GNN usage, researchers have proposed several hybrid and ensemble models with GNN for diverse domains. Catal et al.~\cite{catal2021malware} proposed a GAT model for the transportation domain, and Dvorak et al.~\cite{dvorak2022gnn} for databases. Additionally, Wu et al.~\cite{wu2023deepcatra} proposed GNN paired with LSTM to help capture the structural information.
In Section \ref{subsection:malware_discussion} we discuss the enduring drawbacks with enhancement opportunities.
\vspace{-1mm}

\subsection{Report} \label{subsection:report_summary}
    Reporting is a possible countermeasure against the final CKC phase, Action on Objective (AOO). In AOO phase, the attacker executes its intended goals and often tries to destroy its footprints. To learn the attack TTPs and prevent similar future occurrences, the defensive models require rapid learning updates regarding the attacks. For this, organizations rely on robust reporting platforms such as CVE~\cite{cve} to gather and disclose CTI, such as logs, traffic analysis, preventive measures, and other relevant data, related to novel security incidents and breaches. Cybersecurity analysts at the security operations center (SOC) parse these attack footprints to generate CKG, defensive rules, policies, etc., to update defensive cybersecurity models. Efficient information storage, propagation, and sharing of these security updates is crucial to maintain this pipeline. GNN can influence conducting efficient knowledge propagation, storage, and learning for the ML models from CKGs. Hence, by prompt reporting and efficient knowledge processing through GNN, organizations can collaborate and initiate incident response procedures to equip against novel attacks and prevent similar occurrences with appropriate countermeasures. 

\begin{table*}[!htbp]
\centering
{
    \footnotesize
    {
    \def\arraystretch{1.3}
    \begin{tabular}{ p{3cm}|p{3cm}|p{5cm}|p{3.5cm}  }
    \rowcolor{lightgray!20!}
    %==========================================%
    \hline
    \textbf{Paper Title} & 
    \textbf{Focus/Objective} & 
    \textbf{Contributions} & 
    \textbf{Limitations}\\
    %==========================================%
    \hline
    \rowcolor{cyan!20!}
    \multicolumn{4}{|c|}{\textbf{Fake-Data Detection in Social Media}} \\
    %==========================================%

     %==========================================%
     \hline
      Graph Neural Networks with Continual Learning for Fake News Detection from Social Media~\cite{han2020graph}
     & 
     Propagation-based fake news detection using GNN
     &
      \begin{minipage}[t]{\linewidth}      \begin{itemize}[leftmargin=*]
         \item Used GNN to differentiate between propagation patterns of real and fake news.
         \item Used continual learning to deal with catastrophic forgetting problem.
     \end{itemize}       \vspace{1mm}      \end{minipage} 
     & 
      \begin{minipage}[t]{\linewidth}      \begin{itemize}[leftmargin=*]
         \item Limited to the catastrophic forgetting phenomenon.
         \item GEM and EWC causes increased computation time.
     \end{itemize}       \vspace{1mm}      \end{minipage} \\

     %==========================================%
     \hline
     Propagation-Based Fake News Detection Using Graph Neural Networks with Transformer~\cite{matsumoto2021propagation}
     & 
     Fake news detection using graph transformer network (GTN).
     & 
      \begin{minipage}[t]{\linewidth}      \begin{itemize}[leftmargin=*]
         \item Developed a weight function to enhance the difference propagation pattern between real and fake news.
         \item Improved accuracy by the use of GTN.
     \end{itemize}       \vspace{1mm}      \end{minipage} 
     & 
      \begin{minipage}[t]{\linewidth}      \begin{itemize}[leftmargin=*]
         \item Was compared against two GNN models.
         \item Evaluation was done only using Twitter data.
     \end{itemize}       \vspace{1mm}      \end{minipage} \\

     %==========================================%
     \hline
     Evidence-aware Fake News Detection with Graph Neural Networks~\cite{xu2022evidence}
     & 
     Veracity and evidence-based fake news detection using GNN.
     & 
      \begin{minipage}[t]{\linewidth}      \begin{itemize}[leftmargin=*]
         \item Can learn from complex graph-structured claims and evidence data.
         \item Simple yet effective graph structure learning approach for redundancy mitigation.
     \end{itemize}       \vspace{1mm}      \end{minipage} 
     & 
      \begin{minipage}[t]{\linewidth}      \begin{itemize}[leftmargin=*]
         \item Including other datasets might increase robustness.
         \item Improved performance is required for deployment.
     \end{itemize}       \vspace{1mm}      \end{minipage} \\

     %==========================================%
     \hline
     \rowcolor{cyan!20!}
     \multicolumn{4}{|c|}{\textbf{Cybersecurity Knowledge Graph Improvement}} \\ 
     %==========================================%

     %==========================================%
     \hline
     Cybersecurity Knowledge Graph Improvement with Graph Neural Networks~\cite{dasgupta2021cybersecurity}
     & 
     Generate score for semantic triples in CKG for fake or outdated data detection.
     & 
      \begin{minipage}[t]{\linewidth}      \begin{itemize}[leftmargin=*]
         \item Fake and outdated semantic triple detection using authenticity score (0-1).
         \item The scoring GCN model can work with other triples-generating ML models.
     \end{itemize}       \vspace{1mm}      \end{minipage} 
     & 
      \begin{minipage}[t]{\linewidth}      \begin{itemize}[leftmargin=*]
         \item More research is needed to increase the accuracy.
         \item Requires correct dataset for supervised learning.
     \end{itemize}       \vspace{1mm}      \end{minipage} \\
 
    %==========================================%
    \hline
    \end{tabular}
    }
    \caption{
    Summary of the existing literature on reporting using GNN with scope, contribution, and limitations.}
    \label{table:report_gnns}
}
\vspace{-9mm}
\end{table*}

% Han et al.~\cite{han2020graph} -  Graph Neural Networks with Continual Learning for Fake News Detection from Social Media - 2020
To maintain the authenticity of the data source, Han et al.~\cite{han2020graph} proposed to identify fake data based on the propagation pattern. The authors trained GNN to learn the propagation pattern of tweets among users and classify faked tweets. Due to the vastly differing fake data landscape, ML models do not work well for unseen data using text-based identification methods. Hence, to allow continual learning using incremental training, the authors used Gradient Episodic Memory (GEM) and Elastic Weight Consolidation (EWC) techniques to achieve high detection accuracy for unseen data. Experiments utilizing the Twitter dataset from FakeNewsNet~\cite{shu2020fakenewsnet} evaluated the model's ability to detect fake news with an average accuracy of 75\%.
% Matsumoto et al.~\cite{matsumoto2021propagation} - Propagation-Based Fake News Detection Using Graph Neural Networks with Transformer - 2021
To achieve the same objective with a similar approach, Matsumoto et al.~\cite{matsumoto2021propagation} proposed a novel weight function to identify different propagation patterns between real and fake news in the graph. The authors used a graph transformer network (GTN) model that works using multi-head attention and message-passing mechanisms to search for usable connectivity relations for identification. Evaluation using the previous dataset ~\cite{shu2020fakenewsnet} against two state-of-the-art propagation-based GNN methods demonstrated an improved accuracy of 93\% to 95\%.   
% Xu et al.~\cite{xu2022evidence} - Evidence-aware Fake News Detection with Graph Neural Networks - 2022
Contrary to the propagation pattern of data for fake data identification, Xu et al.~\cite{xu2022evidence} proposed to train GNN models with textual knowledge of evidence-based veracious claims. Multiple claims and evidence are modeled as graphs-structured data to capture the semantic dependency using GNN. Then, using structure learning, long-distance semantic dependencies are captured and mitigated. Finally, the learned model predicts using fine-grained semantic representations. Evaluations using Snopes~\cite{popat2017truth} and PolitiFact~\cite{vlachos2014fact} datasets against baseline state-of-the-art patterns and evidence-based approaches demonstrated its potential and robustness of being used as a plug-in-play approach with other models. 

% Dasgupta et al.~\cite{dasgupta2021cybersecurity} - Cybersecurity Knowledge Graph Improvement with Graph Neural Networks - Dec 2021
Dasgupta et al.~\cite{dasgupta2021cybersecurity} proposed to improve CKG consistency by identifying fake or outdated data by computing a score for each CKG triple using GCN. The authors trained the GNN using manually curated correct data to identify faked or outdated data with a scoring mechanism. Experiments covering diverse malware information from numerous CTI sources and synthetic fake data obtained an F-1 score of 0.975. In Section \ref{subsection:report_discussion}, we address persisting loopholes with improvement areas. Following we provide our complete research information table \ref{table:research_info} with critical research details such as GNN models, classification type, datasets, research timeline, and performance.

\begin{table*}
\footnotesize

% \centering
{
\newcolumntype{x}{>{\columncolor{lightgray!20!}}p{1.5cm}}
\newcolumntype{y}{>{\columncolor{pink!20!}}p{2.4cm}}
\begin{tabular}{x|y|p{2cm}|c|c|p{2.5cm}|c|c}
    \hline
    \rowcolor{cyan!20!}
    \textbf{Prevention Phase} & \textbf{Category} & \textbf{Classification \newline node/edge/graph} & \textbf{Paper} & \textbf{Year} & \textbf{GNN Models / \newline Algorithm} & \textbf{Dataset} & \textbf{Performance} \\  \hline

        % #######################################################################################
        % %=====================================================================================%
        & 
        & graph
        &   ~\cite{li2020adversarial}
        &   2020
        & GAE, GCN, AAE
        &  Rochester
        &  Avg. F1 0.59 \\ \cline{3-8}
        % %==========================================%

        &  
        & node
        &   ~\cite{wang2021privacy} 
        &   2021
        &GCN, GAT, HGCN
        &   ~\cite{mccallum2000automating,giles1998citeseer,sen2008collective}
        &  AUC 95.18 \% \\ \cline{3-8}
        % %==========================================%
        
        &   \multirow{-3}{*}{\parbox{2.5cm}{Adversarial Privacy \\ Preserving}} 
        & graph
        &   ~\cite{liao2021information}
        &   2021
        & GCN, GAT, ChebNet
        &   ~\cite{giles1998citeseer,sen2008collective,konstan1997grouplens,toutanova2015representing}
        &  0.959 $T_{1}$ Perf. \\  \cline{2-8}
        % %==================================================================%
        
        & 
        & graph
        &   ~\cite{wu2021fedgnn} 
        &   2021
        & FedGNN
        &   ~\cite{konstan1997grouplens,jamali2010matrix,ma2011recommender,dror2012yahoo}%, Flixster, Douban, \& YahooMusic
        &  0.989 RMSE \\ \cline{3-8}
        % %==========================================%
        
        &   \multirow{-2}{*}{\parbox{2.5cm}{Federated Learning,\\ Split Learning}}
        & node, graph
        &   ~\cite{shan2021towards}
        &   2021
        & GNN with BP
        &   ~\cite{mccallum2000automating,giles1998citeseer,sen2008collective}
        &   Avg. Acc. 75.33 \%\\  \cline{2-8}
        % %==================================================================%
        
        & 
        & node, graph
        &   ~\cite{hu2022learning} 
        &   2022
        & GCN, DRL, NCL
        &   ~\cite{takac2012data,agarwal2021towards}
        & Acc. 90.62\%  \\ \cline{3-8}
        % %==========================================%
        
            \multirow{-7}{*}{\parbox{1.5cm}{Privacy \\ Maintainence}} 
        &   \multirow{-2}{*}{Others} 
        & graph
        &   ~\cite{wang2023secgnn}
        &   2023
        & GCN
        &   ~\cite{mccallum2000automating,giles1998citeseer,sen2008collective}
        &  Acc. 78.6 \% \\  \hline
        % %=====================================================================================%
        % #######################################################################################
        % %=====================================================================================%
        & 
        &   node, graph
        &   ~\cite{dai2018adversarial} 
        &   2018
        &   RL-S2V, $Q^*$
        &  ~\cite{giles1998citeseer, mccallum2000automating, sen2008collective}
        &  Acc. $\downarrow$: 60\% \\ \cline{3-8}
        % %==========================================%

        &  
        &   node
        &   ~\cite{zugner2019adversarial} 
        &   2019
        &   Meta-learning
        &   ~\cite{giles1998citeseer, mccallum2000automating, adamic2005political}
        &   48\% $\downarrow$ | 5\% Change \\ \cline{3-8}
        % %==========================================%

        &  
        &   node
        &   ~\cite{lin2020:epo} 
        &   2020
        &   Exploratory
        &   ~\cite{giles1998citeseer, mccallum2000automating, bojchevski2017deep} 
        &   10.7\% $\downarrow$ | 3\% Change\\ \cline{3-8}
        % %==========================================%

        &  
        &   edge
        &   ~\cite{lin2020adversarial}
        &   2020
        &   $\Upsilon$-decaying
        &   ~\cite{newman2006finding, watts1998collective, ackland2005mapping}
        &   76.5\% ASR | 25\% Cng. \\ \cline{3-8}
        % %==========================================%

        &  
        &   graph
        &   ~\cite{zhang2021backdoor} 
        &   2021
        &   Backdoor
        &   ~\cite{weber2019anti, wang2017sybilscar, yanardag2015deep}
        &   74\% Avg. ASR \\ \cline{3-8}
        % %==========================================%

        &  
        &   node
        &   ~\cite{zhang2022semantics} 
        &   2022
        &   RL
        &   ~\cite{tan2016artificial, virusshare, virustotal}
        &   Avg. ASR: 93\% \\ \cline{3-8}
        % %==========================================%
        
        &   \multirow{-7}{*}{\parbox{2.5cm}{Adversarial Attack}} 
        &   node
        &   ~\cite{zhou2021hierarchical}
        &   2021
        &   Saliency Map
        &   ~\cite{hamza2019detecting}
        &   Acc. $\downarrow \geq$ 30\% \\  \cline{2-8}
        % %==================================================================%
        
        & 
        &   node, edge
        &   ~\cite{li2020developing} 
        &   2020
        &   Capsule Scheme
        &   ~\cite{ieee30, ieee118}
        &   Recall: 0.98 \\ \cline{3-8}
        % %==========================================%
        
            \multirow{-9}{*}{\parbox{1.5cm}{Adversarial \\ Research}} 
        &   \multirow{-2}{*}{\parbox{2.5cm}{Adversarial Defense}} 
        &   node, edge
        &   ~\cite{boyaci2021graph}
        &   2021
        &   Chebyshev layer
        &   ~\cite{ieee14, ieee118, ieee300}
        &   F1 $\uparrow$ 4\% > CNN\\  \hline
        % %=====================================================================================%
         % #######################################################################################
        % %=====================================================================================%
        & 
        &   node
        &   ~\cite{chaudhary2019anomaly} 
        &   2019
        &   Traditional GNN
        &   ~\cite{enron, twitter}
        &   Accuracy: 98\%\\ \cline{3-8}
        % %==========================================%

        &  \multirow{-2}{*}{\parbox{2.5cm}{Point Anomaly}} 
        &   node
        &   ~\cite{li2021relevance} 
        &   2021
        &   Fusion GCN, GAT
        &   ~\cite{twitter, yelpchi}
        &   Acc.> 79\% | 20\% train \\ \cline{2-8}
        % %==========================================%

        &  
        &   node, graph
        &   ~\cite{ji2021anomaly} 
        &   2021
        &   Deep-GNN
        &   ~\cite{micro_blog, nr}
        % \footnote[1]{{Possible dataset, as exact details were not provided.}}
        &   Accuracy: 95\% \\ \cline{3-8}
        % %==========================================%

        &   \multirow{-2}{*}{\parbox{2.5cm}{Contextual Anomaly}} 
        &   graph
        &   ~\cite{song2021spammer} 
        &   2021
        &   Bayesian Optim.
        &   tagged.com
        &   Accuracy > 96\% \\ \cline{2-8}
        % %==========================================%

        &  
        &   graph, node
        &   ~\cite{wang2021one} 
        &   2021
        &   OCGNN
        &   ~\cite{mccallum2000automating, giles1998citeseer, sen2008collective}
        &   AUC > 0.82 \\ \cline{3-8}
        % %==========================================%

        & 
        &   graph, node
        &   ~\cite{huang2021one} 
        &   2021
        &   TGAT+OCGNN
        &   ~\cite{kumar2019predicting}
        &   AUC >  0.86 \\ \cline{3-8}
        % %==========================================%
        
        &   \multirow{-3}{*}{\parbox{2.5cm}{Collective Anomaly}} 
        &   edge
        &   ~\cite{shi2021multi}
        &   2021
        &   GCN+GAT
        &   ~\cite{openflight, adamic2005political, email_eu_core}
        &   Avg. AUC $\approx$ 0.86 \\  \cline{2-8}
        % %==================================================================%
        
        & 
        &   node
        &   ~\cite{zhao2020gnn} 
        &   2020
        &   Graph Mining Alg.
        &   ~\cite{kumar2016edge, jiang2016catching}
        &   Acc. $\uparrow \approx$ 10\%\\ \cline{3-8}
        % %==========================================%
        
            \multirow{-9}{*}{\parbox{1.5cm}{Anomaly \\ Detection}} 
        &   \multirow{-2}{*}{\parbox{2.5cm}{Others}} 
        &   node, edge
        &   ~\cite{deng2021graph}
        &   2021
        &   Attention
        &   ~\cite{mathur2016swat, ahmed2017wadi}
        &   F1 $\geq$ 0.98 \\  \hline
        % %=====================================================================================%
        % #######################################################################################
        % %=====================================================================================%
        & 
        &   graph
        &   ~\cite{Zhou2019Devign} 
        &   2019
        &    GGNN 
        &  ~\cite{devign_dataset}
        &  Accuracy: 72.26\% \\ \cline{3-8}
        % %==========================================%

        &  
        &   graph
        &   ~\cite{Nguyen2022ReGVD} 
        &   2022
        &   Residual-GGNN
        &   ~\cite{lu2021codexglue}
        &   Accuracy: 63.69\%\\ \cline{3-8}
        % %==========================================%

        &  
        &   graph
        &   ~\cite{Wang2021combining} 
        &   2021
        &   GGNN+stacked GRU
        &   ~\cite{funded}
        &   Accuracy: $\uparrow$ 12.6\%\\ \cline{3-8}
        % %==========================================%

        &  
        &   graph
        &   ~\cite{cao2021bgnn4vd} 
        &   2021
        &   Bidirectional-GGNN
        &   ~\cite{bgnn4vd}
        &   Accuracy: $\uparrow$ 4.9\%\\ \cline{3-8}
        % %==========================================%

        &  
        &   graph
        &   ~\cite{Cheng2021DeepWukong} 
        &   2021
        &   GCN,GAN,k-GNN
        &   ~\cite{DeepWukong}
        &   Accuracy: 97.4\% \\ \cline{3-8}
        % %==========================================%

        &   
        &    graph
        &   ~\cite{tang2023csgvd} 
        &   2023
        &   GNN with M-BFA
        &   ~\cite{lu2021codexglue}
        &   Accuracy: 64.46\%\\ \cline{3-8}
        % %==========================================%
        
        & 
        &   node, graph
        &   ~\cite{rabheru2022hybrid} 
        &   2022
        &   GCN+GRU
        &   ~\cite{sard}
        &   F1 score: 88.12\%\\ \cline{3-8}
        % %==========================================%

        &   \multirow{-8}{*}{\parbox{2.5cm}{Source-code}} 
        &   graph
        &   ~\cite{sefa2022}
        &   2022
        &    GraphSAGE+GCN 
        &   ~\cite{wireshark-dataset}
        &   F1 score: 74.4\%\\  \cline{2-8}
        % %==================================================================%
        
            \multirow{-9}{*}{\parbox{1.5cm}{Vulnerability \\ Detection}} 
        &   \multirow{-1}{*}{\parbox{2.5cm}{Smart Contracts}} 
        &   node, edge
        &   ~\cite{zhuang2020smart}
        &   2020
        &   DR-GCN+TMP
        &   ~\cite{Vntchainwebsite}
        &   Accuracy: 84.48\% \\  \hline
        % %=====================================================================================%
        % #######################################################################################
        % %=====================================================================================%
        & 
        &   edge
        &   ~\cite{lo2022graphsage} 
        &   2021
        &   GraphSAGE
        &   ~\cite{koroniotis2019towards, alsaedi2020ton_iot, sarhan2020netflow}
        &   F1 $\approx$ 0.97\\ \cline{3-8}
        % %==========================================%

        &  
        &   node, edge
        &   ~\cite{chang2021graph} 
        &   2021
        &   GraphSAGE+GAT
        &   ~\cite{cic_daknet, ton-iot, unsw_nb15, cse_cic_ids, sharafaldin2018toward}
        &   F1 $\uparrow$ > 0.01 \\ \cline{3-8}
        % %==========================================%
        
        &   \multirow{-3}{*}{\parbox{2.5cm}{Network Based}} 
        &   edge
        &   ~\cite{pujol2022unveiling}
        &   2022
        &   Traditional GNN
        &   ~\cite{sharafaldin2018toward}
        &   F1 $\approx$ 0.99 \\  \cline{2-8}
        % %==================================================================%
        
        & 
        &   graph
        &   ~\cite{hu2021using} 
        &   2021
        &   Sequence embed.
        &   ~\cite{creech2013semantic}
        &   Acc. $\uparrow \approx$ 2\% \\ \cline{3-8}
        % %==========================================%
        
            \multirow{-5}{*}{\parbox{1.5cm}{Intrusion \\ Detection}} 
        &   \multirow{-2}{*}{\parbox{2.5cm}{Host Based}} 
        &   node
        &   ~\cite{wang2022threatrace}
        &   2022
        &   GraphSAGE
        &   ~\cite{streamspot, unicorn, darpa}
        &   F1 $\approx$ 0.85\\  \hline
        % %=====================================================================================%
        % #######################################################################################
        % %=====================================================================================%
        & 
        &   graph
        &   ~\cite{wang2019heterogeneous} 
        &   2019
        &   HAGNE+Siamese Net
        &   --
        &   Accuracy: 97\% \\ \cline{3-8}
        % %==========================================%

        & 
        &   graph
        &   ~\cite{yan2019classifying} 
        &   2019
        &   DGCNN
        &   ~\cite{ronen2018microsoft, yancfg, virustotal}
        &   F1 $\approx$ 0.97 \& 0.8 \\ \cline{3-8}
        % %==========================================%

        &  
        &   graph
        &   ~\cite{hua2020classifying} 
        &   2020
        &   DGCNN
        &   ~\cite{virusshare}
        &   Accuracy: 96.4\% \\ \cline{3-8}
        % %==========================================%
        
        &   \multirow{-3}{*}{\parbox{2.5cm}{General Malware}} 
        &   graph
        &   ~\cite{li2022intelligent}
        &   2022
        &   GCN+Markov chain
        &   ~\cite{virustotal, virusshare}
        &   Accuracy: 98.3\% \\  \cline{2-8}
        % %==================================================================%
        
        & 
        &   graph
        &   ~\cite{feng2021android} 
        &   2021
        &   GRL
        &   ~\cite{arp2014drebin, allix2016androzoo}
        &   Accuracy: 97.1\% \\ \cline{3-8}
        % %==========================================%
        
        &   
        &   edge, graph
        &   ~\cite{busch2021nf}
        &   2021
        &   OCGNN
        &   ~\cite{lashkari2018toward}
        &   Accuracy: 95\%  \\  \cline{3-8}
        % %==================================================================%
        
        &   \multirow{-3}{*}{\parbox{2.5cm}{Android Malwares}}
        &   graph
        &   ~\cite{weng2022graph}
        &   2022
        &   GNN+JK
        &   ~\cite{freitas2020large, arp2014drebin}
        &   Acc. $\uparrow \approx$ 8\% \\ \cline{2-8}
        % %==========================================%
        
            \multirow{-8}{*}{\parbox{1.5cm}{Malware \\ Detection}} 
        &   \multirow{-1}{*}{\parbox{2.5cm}{IoT}} 
        &   graph
        &   ~\cite{li2021cross}
        &   2021
        &   GraphSAGE
        &   ~\cite{virusshare}
        &   Accuracy: 99.61\% \\  \hline
        % %=====================================================================================%
        % #######################################################################################
        % %=====================================================================================%

        &  
        &   graph
        &   ~\cite{han2020graph} 
        &   2020
        &   DiffPool
        &   ~\cite{shu2020fakenewsnet}
        &   Accuracy: 75\%\\ \cline{3-8}
        % %==========================================%

        &  
        &   edge, graph
        &   ~\cite{matsumoto2021propagation} 
        &   2021
        &   GTN
        &   ~\cite{shu2020fakenewsnet}
        &   Accuracy: 95\% \\ \cline{3-8}
        % %==========================================%
        
        &   \multirow{-3}{*}{\parbox{2.5cm}{Fake Data}} 
        &   node, graph
        &   ~\cite{xu2022evidence}
        &   2022
        &   Structure Learning
        &   ~\cite{popat2017truth, vlachos2014fact}
        &   F1 $\approx$ 0.7 \& 0.8\\  \cline{2-8}
        % %==================================================================%
        
            \multirow{-4}{*}{\parbox{1.5cm}{Threat \\ Report}} 
        &   \multirow{-1}{*}{\parbox{2.5cm}{Knowledge Graph}} 
        &   node, edge
        &   ~\cite{dasgupta2021cybersecurity}
        &   2021
        &   GCN
        &   --
        &   F1: 0.975 \\  \hline
        % %=====================================================================================%
        
\end{tabular}}
\vspace{1mm}
\caption{Survey research information summary Table containing all articles following our taxonomy. Notations are following. [ASR: Attack Success Rate], [$\uparrow$: Increase], [$\downarrow$: Decrease], [$|$: At], [Acc. : Accuracy], [Cng. : Change], [Perf. : Performance]}\label{table:research_info}
\end{table*}

\section{Discussion}  \label{section:discussion} %7-8 Pages
In this section, we will discuss the ongoing research limitations and potential future research directions. We address the improvement scopes based on our summarized research findings in Section \ref{section:gnn_in_cyber}. 

    \subsection{Privacy Maintenance} \label{subsection:privacy_discussion}
    Privacy indicates how private data, such as activity feed, system data, and security model parameters, can be protected to prevent it from being leaked. In an effort to prevent leakage, privacy-preserving GNNs are being developed. In the body of literature, these GNNs are categorized into various categories, such as adversarial privacy preservation~\cite{li2020adversarial, wang2021privacy, liao2021information}, federated learning~\cite{ wu2021fedgnn}, differential privacy \& split Learning~\cite{ shan2021towards}, and others~\cite{hu2022learning, wang2023secgnn}.
    \vspace{-5mm}
    
    \begin{figure*}[ht]
      \centering
      \includegraphics[width=0.95\textwidth]{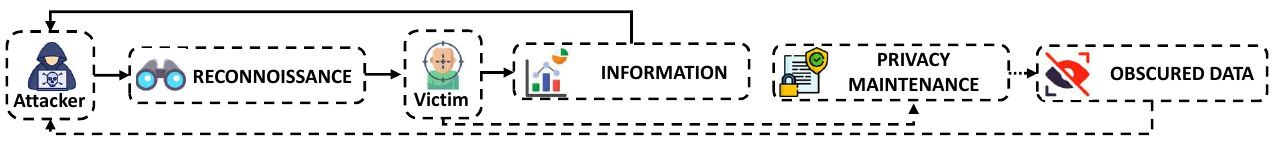}
      \vspace{-3mm}
      \caption{Privacy Maintenance as a security measure against reconnaissance to obfuscate victim information.} 
      \label{Fig:privacy_diag}
      \vspace{-3mm}
    \end{figure*}
    \vspace{-2mm}

    Researchers are currently working on metrics to evaluate the effectiveness of privacy-preserving GNNs. These models, which are based on adversarial privacy-preserving techniques, such as those presented in~\cite{li2020adversarial, wang2021privacy, liao2021information}, focus on obtaining high accuracy in predicting link and utility attributes while protecting privacy. Li et al.~\cite{li2020adversarial}, for example, achieved superior privacy protection and utility retention, achieving an average Macro-F1 score of 0.549 on private attributes. On the other hand, despite effectively maintaining a high level of privacy, some authors observed a drop in link prediction accuracy when applying node privacy protection~\cite{wang2021privacy}. In a similar vein, Liao et al.~\cite{liao2021information} also reported a slight loss in task performance; nevertheless, their models demonstrated robustness against inference attacks. These approaches, however, fall short of effectively addressing how to achieve an ideal compromise between prediction performance and privacy protection. More research is required to thoroughly investigate this essential feature.
    Additionally, when employing privacy-preserving approaches such as federated learning and differential privacy~\cite{wu2021fedgnn}, it is critical to carefully examine hyperparameters in order to strike the right balance between model performance and privacy protection. If the privacy budget is set too low, it can have a negative impact on the accuracy of model gradients. Conversely, work that emphasizes privacy preservation in a distributed setting, such as Split Learning~\cite{shan2021towards}, tends to ignore the formal notion of differential privacy, leading to considerably weaker privacy guarantees. Addressing these challenges and improving the overall privacy-performance trade-off is critical for achieving robust and effective privacy-preserving GNNs.
    In a separate line of work, Hu et al.~\cite{hu2022learning} have made an effort to address the problem of privacy leakage concerning private user data by disentangling node features and enforcing orthogonality within a relevant space. Despite the fact that elements may be orthogonal, it is important to recognize that statistical correlations might exist that could potentially lead to privacy breaches. Similarly, Wang et al.~\cite{wang2023secgnn} have presented a solution for privacy-preserving GNNs in an outsourced scenario. However, the approach might not offer sufficient resilience against sophisticated attacks. 
    % \vspace{-2mm}

    \subsection{Research}  \label{subsection:research_discussion}
        GNNs have emerged as powerful tools for analyzing structured data, such as social networks, CFGs, DFGs, CKGs, etc. However, recent studies have highlighted the vulnerability of GNNs to adversarial attacks, where subtle perturbations to the input graph can lead to malicious node, edge, and graph predictions. Adversarial research has primarily focused on injecting continuous perturbations in the target graph without breaking the combinatorial structure~\cite{dai2018adversarial, zugner2019adversarial, zhang2021backdoor, zhou2021hierarchical, lin2020:epo, lin2020adversarial}. To understand the research landscape, we categorized it into the following:
        % \vspace{-3mm}
        % To attain such, Dai et al.~\cite{dai2018adversarial} developed reinforcement learning, genetic algorithm, and gradient descent-based approaches, whereas Zügner et al.~\cite{zugner2019adversarial} proposed meta-learning. Zhang et al.~\cite{zhang2021backdoor} developed a backdoor attack for black-box scenarios when the attacker does not have any knowledge about the testing graph, which was further enhanced by Zhou et al.~\cite{zhou2021hierarchical} by introducing a saliency map technique on shadow GNN model. On the other hand, exploratory greedy~\cite{lin2020:epo} and greedy heuristic~\cite{lin2020adversarial} approaches were tested to achieve optimal results that might not get attained using gradient descent.

            \subsubsection{Performance \& Transferability}
                The impact of any developmental research is measured by comparing its performance matrix, which includes accuracy, robustness, transferability, etc. From an accuracy point of view, reinforcement learning, genetic algorithm, and gradient descent-based approaches~\cite{dai2018adversarial} were able to achieve 40\% to 60\% misclassification, and 48\% misclassification was achieved with 5\% perturbed edges by Zügner et al.~\cite{zugner2019adversarial}. Another similar performance drop compared to state-of-the-art models was noted in the works of Lin et al.~\cite{lin2020:epo}. Contrary to black-box scenarios, misclassification accuracy in white-box settings was relatively higher, around 90\%~\cite{lin2020adversarial} to 100\%~\cite{zhang2021backdoor}. Using such approaches misclassification accuracy was observed consistent throughout, justifying the severity. Apart from a consistent misclassification trend, another notable factor was transferability. Most of the attacks are model-independent and thus can be transferred and applied to other models. For example, the attack developed by Zhang et al.~\cite{zhang2022semantics} can be operated on any GNN model that utilizes sequential data. Hence, the consistency and robustness of the adversarial attacks pose a severe threat to the employed GNN models.
                \vspace{-3mm}

                \begin{figure*}[ht]
                  \centering
                  \includegraphics[width=0.9\textwidth]{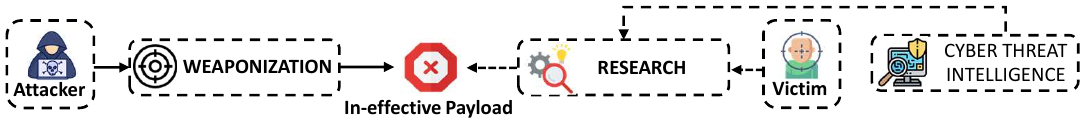}
                  \vspace{-3mm}
                  \caption{Research as a proactive measure against weaponization to prepare against possible attack.} 
                  \label{Fig:research_diag}
                  \vspace{-3mm}
                \end{figure*}
                \vspace{-3mm}
                
            \subsubsection{Defence mechanism \& mitigation}
                Compared to the active research on developing adversarial attacks on GNN models, there are only a few in defense~\cite{li2020developing, boyaci2021graph}. Some methods include incorporating adversarial samples in the training data~\cite{dai2018adversarial} or randomized smoothing~\cite{zhang2021backdoor} as a primary defense strategy. Furthermore, contrary to adversarial attacks, defensive approaches are not equally transferable and are mostly evaluated over some synthesized datasets. To address this, researchers attempted to increase the explainability of GNN models~\cite{he2022illuminati, ying2019gnnexplainer}. However, such models are scope-dependent. From a defensive standpoint, robust explanations are needed. Therefore, immediate attention to robust explainability and defensive GNN research is needed. 

        % Adversarial attacks pose a significant threat to the integrity and performance of GNNs. By actively focusing on explainability~\cite{neupane2022explainable}, we can deepen our understanding of the vulnerabilities of GNN models and develop robust defense mechanisms to mitigate the impact of adversarial attacks. The existing research has the potential to advance the field of graph-based machine learning and contribute to the development of secure cyber operations using GNN.

    \subsection{Anomaly Detection} \label{subsection:anomaly_discussion}
        Anomaly detection is a critical task in cyber defense to prevent the delivery of malicious payloads and ensure end-to-end security with significant impact in various domains, such as finance, healthcare, industrial systems, etc. Traditional anomaly detection methods often struggle to capture complex network patterns and dependencies present as graph-structured data. GNNs have shown promise in modeling and analyzing graph data, making them an attractive candidate for anomaly detection tasks. In a generic approach, by learning the standard behavioral relation between nodes, GNN is used to classify the deviant. Apart from generic GNN approaches, researchers have also proposed a few indirect ones to enhance GNNs' functionality for anomaly detection~\cite{zhao2020gnn, deng2021graph}. Other than conventional scenarios, researchers also addressed a few different long-standing ones such as DoS attacks~\cite{chakraborty2023capow} through anomaly detection. We evaluate the research landscape and potential improvement scopes from an implementable point of view through the following categories.
        \vspace{-3mm}
        
            \begin{figure*}[ht]
              \centering
              \includegraphics[width=0.85\textwidth]{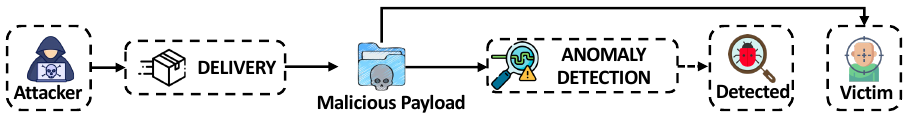}
              \vspace{-3mm}
              \caption{Anomaly Detection to detect malicious payload against weaponization to prepare against possible attack.} 
              \label{Fig:anomaly_diag}
              \vspace{-5mm}
            \end{figure*}
            % \vspace{-3mm}

            \subsubsection{Scalability \& Cost}
            Cyber defense systems often handle large-scale and dynamic graph data, posing challenges in terms of scalability and efficiency for GNN-based anomaly detection. Even with achieving impressive accuracy in limited testing data, models often struggle to handle large chunks of real-time data with significant accuracy~\cite{chaudhary2019anomaly, li2021relevance, song2021spammer, wang2021one}. One of the root causes of such occurrences stems from higher dependency among various interconnected models designed to handle specific scenarios~\cite{ji2021anomaly, wu2021graph}. Hence, developing scalable algorithms using modular architectures that can process and analyze large graphs from diverse aspects in real-time is crucial for deployment. Fortunately, from hardware standpoint, suppliers have started developing chips to support such extensive computation~\cite{gnn_intel_hardware}. On the other hand, few models require pre-curated data for training~\cite{huang2021one}, increasing costs. Therefore, robust yet self-learning aspects during model development demand significant attention. 
            
            \subsubsection{Interpretability \& Explainability}
            Anomalies in cyber systems can exhibit diverse patterns, including rare events, stealthy attacks, and evolving behaviors. Additionally, existing models generate extensive alerts that cybersecurity analysts are not able to verify on time, known as threat alert fatigue problem~\cite{he2022illuminati}. Ensuring that GNNs only capture and identify genuine anomalies, requires careful modeling and innovative techniques. Researchers have tried to improve the fine-tuning functions~\cite{zhao2020gnn, deng2021graph} to improve accuracy– instead of direct model improvement. Leveraging GNN explainability from different aspects~\cite{ying2019gnnexplainer, yuan2021explainability, he2022illuminati} will facilitate model improvement research in a directed manner. Hence, contextualized and explainability-driven GNN model development will correctly interpret underlying anomalous patterns and reduce analyst intervention with low false-positives. 

       % \st{ Anomaly detection in cyber defense using GNNs holds significant potential for improving the detection of complex and interconnected occurrences. Despite the challenges, the capabilities of GNN can enhance accuracy and effectiveness. By addressing the challenges discussed above, GNN-based anomaly detection in cyber defense can pave the way for proactive defensive cyber operations in the digital realm.} \hl{we dont need this paragraph}

    \subsection{Vulnerability Detection} \label{subsection:vulnerability_discussion}
    GNN-driven vulnerability detection complements human expert-based vulnerability finding and identification of potential weaknesses and security risks in complex source code. Evidently, GNN has shown effective ways to capture the relationships among code elements, such as functions, and variables, enabling them to identify various types of vulnerabilities, such as buffer overflows, injection attacks, code injections, etc. More specialized applications of vulnerability detection exist in blockchain-based smart contract source code where considerable research has been performed~\cite{zhuang2020smart, Cai2023combine,liu2021combining,zeng2022ethergis}. Based on recent progress, we will outline the possible directions.
    \vspace{-4mm}
    
        \begin{figure*}[ht]
          \centering
          \includegraphics[width=0.9\textwidth]{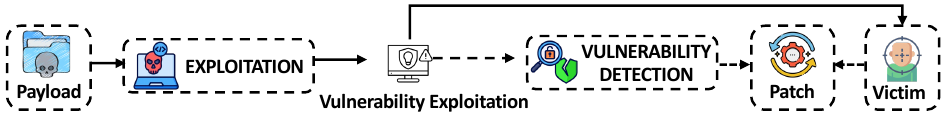}
          \vspace{-3mm}
          \caption{Vulnerability detection to detect vulnerabilities and prevent exploitation by malicious payload by developing patches.} 
          \label{Fig:vulnerability_diag}
        \end{figure*}
        \vspace{-5mm}

    \subsubsection{Secure Coding \& Development.} 
    The utilization of advanced Integrated Development Environments (IDEs) can wield a significant influence in thwarting the introduction of vulnerabilities within the source code. These IDEs offer supportive tools that adeptly steer developers toward crafting secure code. A variety of static code analysis tools offer capabilities tailored to identify vulnerabilities specific to programming languages. This is achieved by systematically considering edge cases, boundary conditions, etc., preemptively. However, to fully harness the capabilities of GNN-based techniques, it is crucial to explore their generalization potential across diverse domains such as software, network, and IoT. In a similar vein, enhancing the interpretability of GNN-based vulnerability detection~\cite{he2022illuminati} systems will provide the means to explain the reasoning behind the predictions made by these systems, which in turn, have greater trust impact, especially for human-in-the-loop systems. Additionally, the practice of efficient graph generation~\cite{Nguyen2022ReGVD, Zhou2019Devign} and code refactoring proposed by several research~\cite{Desai2021graph,suneja2020learning, LeClair2020Improved} can utilize modern large language models (LLMs)~\cite{jain2023code} to alleviate vulnerability introduction further. Another intriguing avenue for research involves investigating the seamless integration of GNNs with existing reverse engineering tools, such as IDA Pro~\cite{idapro} and GIDRA~\cite{ghidra}, could potentially be synergized with GNN-based approaches to enhance the overall efficacy of vulnerability management practices.
    
    \subsubsection{Dynamic \& Heterogenous Benchmarking.} The landscape of source code management is heterogeneous (i.e. includes various programming languages, libraries, frameworks) and dynamic (i.e. addition of new code to patch source code or addition of new logic). Since GNNs trained on one specific codebase might struggle to generalize effectively to different codebases that possess varying coding styles, languages, or domain-specific characteristics, the process of adapting GNNs from one project to another necessitates thorough retraining or fine-tuning. Hence, vulnerability detection in a heterogeneous and dynamic coding landscape poses unique challenges as current literature lacks in providing a diverse benchmark dataset. Many prior research~\cite{Nguyen2022ReGVD, Cheng2021DeepWukong, liu2021combining, tang2023csgvd} is conducted on well-known datasets, such as SARD~\cite{sard}, CodeXGLUE~\cite{lu2021codexglue}, etc. Comparing the results from one with another may not hold due to the varying specifications of each of these datasets, hence requiring further investigation.

    % The potential of GNN-based vulnerability management in source code has been demonstrated in a recent study by Cheng et al.~\cite{Cheng2021DeepWukong}, with an impressive accuracy of 97.4\% and an F1-score of 95.6\%. To fully harness the capabilities of GNN-based techniques, it is crucial to explore their generalization potential across diverse domains such as software, network, and IoT. In a similar vein, enhancing the interpretability of GNN-based vulnerability management systems will provide the means to explain the reasoning behind the predictions made by these systems, which in turn, have greater trust impact, especially for human-in-the-loop systems. Since GNNs trained on one specific codebase might struggle to generalize effectively to different codebases that possess varying coding styles, languages, or domain-specific characteristics, the process of adapting a GNN from one project to another necessitates thorough retraining or fine-tuning. Moreover, an intriguing avenue for research involves investigating the seamless integration of GNNs with existing vulnerability assessment tools, such as IDA Pro and GIDRA \hl{cite these}, could potentially be synergized with GNN-based approaches to enhance the overall efficacy of vulnerability management practices.

    \subsection{Intrusion Detection} \label{subsection:intrusion_discussion}
        Intrusion detection plays a critical role in safeguarding our digital systems by identifying unauthorized or malicious activities within computer networks. As the complexity and sophistication of attacks continue to evolve, traditional intrusion detection methods often struggle to capture the complex and dynamic relationships among processes and network entities, making them susceptible to sophisticated attacks. For such reasons, GNNs have recently gained attention as a potential solution for intrusion detection~\cite{bilot2023graph}. By learning the network and system behavior, GNN models can identify occurring internal threats. However, there are still challenges requiring attention. Additionally, there have been very few works on Intrusion Response Systems (IRS) alongside detection– even when rapid response is necessary to complement efficient detection. For example, risk assessment, prioritization, and parallelization are critical areas that need to be addressed with effective IDS research~\cite{anwar2017intrusion}. Following, we'll discuss the research landscape with challenges and future directions.
        \vspace{-3mm}

            \begin{figure*}[h]
              \centering
              \includegraphics[width=0.8\textwidth]{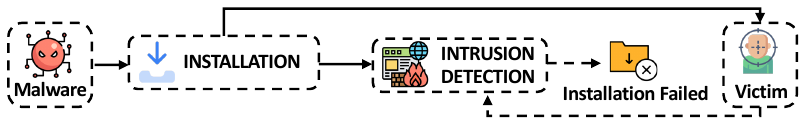}
              \vspace{-3mm}
              \caption{Intrusion detection to prevent malware installation and initial foothold establishment.} 
              \label{Fig:intrusion_diag}
            \end{figure*}
            \vspace{-5mm}

            \subsubsection{Dynamic Environments \& Real-time Detection}
            Intrusion detection models need to handle dynamic network environments, adapt to changing network topologies, and efficiently process streaming data for timely detection and response. Even though GNN models have significantly improved detection accuracy, they still lack to adapt to dynamic environments~\cite{lo2022graphsage, chang2021graph}. Insufficient open-source authentic data replicating the dynamic behavior remains a crucial blocker. Furthermore, intrusion detection requires real-time monitoring and response to rapidly evolving threats. For such, researchers propose advanced and complex GNN models that proportionally increase computation time~\cite{hu2021using}. On the other hand, performance is a critical factor in responding to active threats. Hence, computationally efficient yet robust hybrid model development remains open to possibilities. 
    
            \subsubsection{Trust \& Explainability}
            Trust plays a crucial factor in effective IDS deployment, as it needs full system access to work. Adversaries often target employed IDS systems as a part of reconnaissance activity~\cite{corona2013adversarial}. Furthermore, advanced GNN model predictions have limited explainibility by security analysts~\cite{lo2022graphsage}– few models require either authentic training data~\cite{wang2022threatrace} or synthesized data~\cite{chang2021graph} to train. Some of the models do address explainability without prior knowledge~\cite{ying2019gnnexplainer, yuan2021explainability}, but fail to consider all attributes equally. Hence, data poisoning or adversarial attacks remain a significant threat to ensuring secure deployment and remain an active research direction. Moreover, multi-facet explainability~\cite{neupane2022explainable} of GNN-based intrusion detection models can provide insights into the decision-making process, enhance trust, and facilitate collaboration between distributed detection systems.

        % \st{Intrusion detection in cyber defense using GNN has the potential to counter adversarial activities and detect sophisticated network and system-level intrusions. By leveraging GNN capabilities, IDS can achieve improved yet efficient detection and response. Addressing challenges related to robustness, interpretability, and trust factors will be crucial for practical deployment.} \hl{feels like this is unnecessary and can this be mixed with "Interpretability and Explainability" in section above and have "Trust" component there?}

    \subsection{Malware Detection} \label{subsection:malware_discussion}
        Malware detection aims to identify and mitigate malicious software that poses a threat to computer systems and networks. GNNs have emerged as a promising technique for malware detection, leveraging the inherent graph structure of malware to capture intricate process relationships and dependencies. For example, most research focused on utilizing the malware application CFG~\cite{hua2020classifying, feng2021android, wu2023deepcatra} to map and learn the application's behavior. Additionally, a few utilized network flow graphs (NFG)~\cite{busch2021nf, zhou2020automating} to learn the network communication pattern. However, there are still improvement scopes utilizing GNNs, discussed following.

            \subsubsection{Dynamicity \& Scalability}
            Malware detection systems operate on vast collections of malware samples, requiring efficient processing and analysis. Developing scalable architectures and algorithms that can handle large-scale system call graphs with real-time detection is crucial for deployment. To allow GNN process large graphs, researchers directed to incorporate advanced ML approaches like attention mechanism, JK, etc.,~\cite{wang2019heterogeneous, yan2019classifying, hua2020classifying, weng2022graph}. However, these often consume high training time~\cite{yan2019classifying} and become scenario-specific and less robust~\cite{yan2019classifying, li2022intelligent, feng2021android, weng2022graph, li2021cross} to adapt with evolving malwares. Additionally, low computational devices like mobile or IoT cannot execute such heavy graph extraction and classifier models~\cite{mitra2023survey}. Hence, to improve performance with dynamic adaptability, more research is needed for powerful yet efficient graph extractors and ML classifiers.
            \vspace{-3mm}

             \begin{figure*}[ht]
              \centering
              \includegraphics[width=0.9\textwidth]{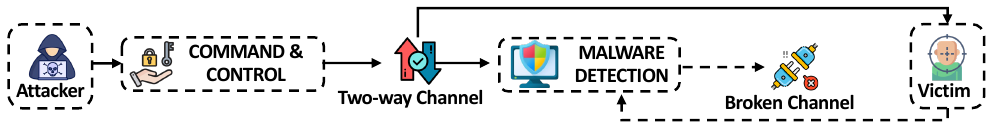}
              \vspace{-4mm}
              \caption{Malware detection and removal disrupts two-way communication channels and prevents persistent control.} 
              \label{Fig:malware_diag}
            \end{figure*}
            \vspace{-4mm}
    
            \subsubsection{Relation Modeling \& Knowledge Sharing}
            GNNs' capability to leverage heterogeneous graphical data to learn relationships between system-process interactions, such as function calls, control flow, or data dependencies, allows effective capture of system behavior information caused by malware samples and spot sophisticated obfuscation techniques by polymorphic malware variants that would go undetected with signature-based approaches. However, GNN being a relatively novel research paradigm, even advanced models often suffer from completely learning the complex graph obfuscation techniques like NOP insertion, subroutine reordering, etc,~\cite{yan2019classifying, feng2021android, mitra2023survey}. A possible solution to allow different GNN and ML models learn diverse relationships on the fly, is through knowledge sharing. Encouraging collaboration and knowledge sharing among the deployed models can foster the development of standardized cybersecurity knowledge graphs~\cite{mitra2021combating}, benchmarks, and evaluation metrics. Such efforts will facilitate addressing diverse behavioral aspects in real-time learning with collective advancements.

       % \st{ The use of GNNs for malware detection in cyber defense offers significant potential to improve the accuracy and effectiveness of detecting sophisticated and evolving malware threats. By leveraging the benefits of GNNs, malware detection systems can enhance their detection capabilities. Addressing challenges related to scalability, performance,  and collaborations will be crucial for industry-wide practical deployment of GNN-based malware detection systems.} \hl{can skip this part and have 2 liner to introduce next section..dont need to reiterate}

    \subsection{Report} \label{subsection:report_discussion}
    Reporting and knowledge sharing play vital roles in cyber defense, as they facilitate the dissemination of actionable intelligence, enable collaboration among cybersecurity professionals, and enhance the collective understanding of emerging threats. GNNs have demonstrated their effectiveness in modeling and analyzing graph-structured data, making them a potential tool for processing knowledge graphs and facilitating learned knowledge for defensive cyber operations. However, the usage of GNN for knowledge sharing has not been given much attention for a long time. Recently a few researchers have utilized the benefit of GNN in fake data detection~\cite{dasgupta2021cybersecurity} and knowledge sharing using federated learning~\cite{he2021fedgraphnn, mei2019sgnn}. Following, we discuss the benefits, challenges, and future directions.
    \vspace{-3mm}

        \begin{figure*}[ht]
          \centering
          \includegraphics[width=0.9\textwidth]{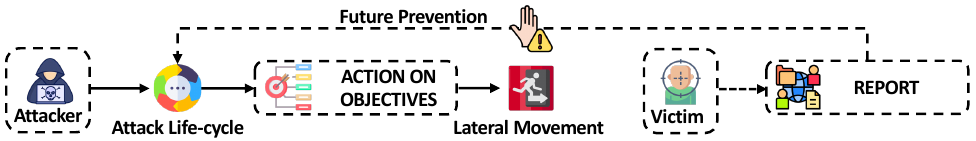}
          \vspace{-3mm}
          \caption{Reporting to disclose attack information and prevent similar attack occurrences in future.} 
          \label{Fig:report_diag}
        \end{figure*}
        \vspace{-5mm}

        \subsubsection{Data Integration \& Collaboration}
        CTI is often accumulated from various sources, including network logs, threat intelligence feeds, security incident reports, and vulnerability databases. Integrating and fusing these heterogeneous data sources into a unified graph representation while maintaining authenticity presents a significant challenge. For authenticity, researchers proposed to use GNN for fake-data identification~\cite{han2020graph, matsumoto2021propagation, xu2022evidence}. However, research on data integration from multiple sources is still left behind. The absence of evenly distributed open-source datasets~\cite{matsumoto2021propagation, xu2022evidence}, the complexity of GNN models~\cite{han2020graph}, and high computation costs~\cite{dasgupta2021cybersecurity} act as blockers and remain an open area of research. To mitigate this, researchers can leverage transfer learning techniques with GNN to share acquired knowledge among related domains or datasets ~\cite{han2021adaptive}.
        \vspace{-1mm}

        \subsubsection{Privacy \& Security}
        Ensuring confidentiality, integrity, and controlled access to shared knowledge while preserving individual and organizational privacy is crucial for successful reporting and knowledge-sharing. Due to the distribution of sensitive cyber defense knowledge across various sources, adversaries target such communications to steal information. Transfer of sensitive cybersecurity data and model gradients adhering to privacy and security concerns remains a significant challenge. In generic research, Mei et al. proposed to safeguard transferred knowledge by hiding the graph structure~\cite{mei2019sgnn}. However, in cybersecurity, limited attention is given to secure knowledge-sharing among employed defensive ML models, and requires immediate attention.

    % Utilizing GNNs for reporting and knowledge sharing in cyber defense holds significant potential in enhancing the understanding, collaboration, and response to emerging threats. By leveraging the benefits of GNNs, such as graph representation, contextual insights, knowledge extraction, and transfer learning, reporting systems can provide comprehensive and actionable intelligence. Continued research and innovation in this field will contribute to more effective reporting, collaboration, and proactive cyber defense strategies.

\section{Conclusion}  \label{section:conclusion} % 1 page (Conclusion & Acknowledgement)
As our world becomes more interconnected and reliant on digital systems, the imperative to safeguard our information and digital infrastructure has never been greater. The relentless evolution of cyber threats has compelled us to explore innovative solutions, and among these, GNN has emerged as a beacon of promise. This paper has delved into the applications of GNNs in disrupting each stage of the Cyber Kill Chain, a well-established attack life-cycle framework for understanding cyberattacks. By addressing each phase and elucidating how GNNs bolster defensive cyber operations, we have underscored the potential of this technology. However, it is crucial to acknowledge that the cyber landscape is dynamic and ever-changing, and GNNs also need to adapt to the change. Undoubtedly there are open research areas and avenues for further improvements. This ongoing exploration is vital to stay ahead of increasingly sophisticated adversaries. With this essence, this paper underscores the contributions and further research directions for GNN to defend against cyber threats effectively.

\begin{acks}
This work was supported by PATENT Lab (Predictive Analytics and TEchnology iNTegration Laboratory) at the Department of Computer Science and Engineering, Mississippi State University. We also acknowledge Kyanie Waters and La’Andrea Gates for their contributions.
\end{acks}

%refs 4-5 pages 
\bibliographystyle{unsrt}
% % \bibliographystyle{ACM-Reference-Format}
\bibliography{sample-base}

\end{document}